\documentclass[conference]{IEEEtran}
\usepackage{booktabs}   
\usepackage{tabularx} 
\usepackage{multirow}  
\usepackage{array}  
\usepackage{makecell} %
\usepackage{pifont}
\usepackage{balance}

\def\BibTeX{{\rm B\kern-.05em{\sc i\kern-.025em b}\kern-.08em
T\kern-.1667em\lower.7ex\hbox{E}\kern-.125emX}}

\newcommand{\eg}{{\em e.g.,}\xspace}
\newcommand{\ie}{{\em i.e.,}\xspace}

\newcommand{\PP}[1]{\vspace{2px}\noindent{\bf#1.}}

\newcommand*\WC[1]{%
	\begin{tikzpicture}[baseline=(C.base)]
		\node[draw,circle,inner sep=0.2pt](C) {#1};
\end{tikzpicture}}

\newcommand{\cc}[1]{\mbox{\smaller[0.5]\texttt{#1}}}

\newcommand{\UP}[1]{\textcolor{black}{#1}} %

\usepackage{cite}
\usepackage{amsmath,amssymb,amsfonts}
\usepackage{textcomp}
\usepackage{tabularx}
\usepackage{caption}
\usepackage{xspace}
\usepackage{enumitem}
\usepackage{relsize}
\usepackage{graphicx}
\usepackage{booktabs}
\usepackage{url}
\usepackage{amsmath}
\usepackage{array}
\usepackage{listings}
\usepackage{algorithm}
\usepackage{algpseudocode}
\usepackage{color}
\usepackage{tikz}
\usetikzlibrary{positioning,arrows.meta,shapes.geometric,fit,backgrounds,calc}
\usepackage{pifont}
\usepackage{ragged2e}
\usepackage[utf8]{inputenc} 
\usepackage[T1]{fontenc}
\usepackage{subcaption}
\usepackage{booktabs, multirow} %
\usepackage{soul}%
\usepackage{xcolor,colortbl} %
\usepackage{changepage,threeparttable} %
\usepackage{longtable}  
\usepackage{tocbibind}
\usepackage[english]{babel}
\usepackage{hyperref}

\usepackage[nameinlink]{cleveref}
\crefformat{section}{\S#2#1#3}

\usepackage{booktabs}
\usepackage{multirow}
\usepackage{siunitx}
\usepackage{stfloats}
\usepackage{eso-pic}

\usepackage[most]{tcolorbox}
\usepackage{lipsum}

\ifdefined\AddToHook %
  \AddToHook{env/table/begin}{}
  \AddToHook{env/table*/begin}{}
\fi

\newcounter{counter}
\newtcolorbox{takeawaybox}[1][]{
before upper = {\stepcounter{counter}},
  colback=white,
  colframe=black!40,       %
  fonttitle=\bfseries,
  title=Takeaway~\#\arabic{counter},
  coltitle=black,
  boxrule=0.5pt,
  arc=3pt,
  left=6pt,
  right=6pt,
  top=6pt,
  bottom=6pt,
}

\providecommand{\cc}[1]{\mbox{\smaller[0.5]\texttt{#1}}}

\providecommand{\eg}{{\em e.g.,}\xspace}
\providecommand{\ie}{{\em i.e.,}\xspace}

\providecommand{\PP}[1]{}
\renewcommand{\PP}[1]{%
  \par\smallskip\noindent\textbf{#1.}\enspace\ignorespaces%
}

\providecommand*\WC[1]{%
	\begin{tikzpicture}[baseline=(C.base)]
		\node[draw,circle,inner sep=0.2pt](C) {#1};
        
\end{tikzpicture}}

\providecommand{\iconnum}{%
    \tikz[baseline=0pt,x=1mm,y=1mm]{%
        \useasboundingbox (0,0) rectangle (2.4,2.4);
        \fill[black!80] (0.40,0.2) rectangle (0.85,0.88);
        \fill[black!80] (0.98,0.2) rectangle (1.43,1.38);
        \fill[black!80] (1.56,0.2) rectangle (2.01,2.08);
    }%
}
\providecommand{\iconseq}{%
    \begin{tikzpicture}[baseline=-0.5ex]
      \draw (0.04,0.06) -- (0,0) -- (0.04,-0.06);
      \fill (0.06,0) circle (0.015);
      \fill (0.11,0) circle (0.015);
      \draw (0.13,0.06) -- (0.17,0) -- (0.13,-0.06);
    \end{tikzpicture}
}

\providecommand{\icongraph}{%
    \begin{tikzpicture}[baseline=-0.9ex]
        \draw (0,0) circle (0.025);
        \draw (0.09,0) circle (0.025);
        \draw (0.09,-0.09) circle (0.025);
        \draw (0.025,0) -- (0.055,0);
        \draw (0.09,-0.025) -- (0.09,-0.055);
    \end{tikzpicture}
}
\providecommand{\iconimg}{%
    \tikz[baseline=0pt,x=1mm,y=1mm]{%
        \useasboundingbox (0,0) rectangle (2.4,2.4);
        \fill[black!80] (0.35,1.20) rectangle (1.20,2.05);
        \fill[black!20] (1.20,1.20) rectangle (2.05,2.05);
        \fill[black!20] (0.35,0.35) rectangle (1.20,1.20);
        \fill[black!80] (1.20,0.35) rectangle (2.05,1.20);
        \draw[black!70,line width=0.12pt] (0.35,0.35) rectangle (2.05,2.05);
        \draw[black!70,line width=0.10pt] (1.20,0.35) -- (1.20,2.05);
        \draw[black!70,line width=0.10pt] (0.35,1.20) -- (2.05,1.20);
    }%
}

\providecommand{\iconA}{%
    \textcircled{a}\xspace
}

\providecommand{\iconS}{%
    \textcircled{s}\xspace
}

\providecommand{\iconR}{%
    \textcircled{r}\xspace
}

\providecommand{\iconD}{%
    \textcircled{d}\xspace
}

\newcolumntype{Y}{>{\RaggedRight\arraybackslash}X}

\makeatletter  %
\@ifundefined{ctw}{}{}\@ifundefined{stuw}{}{}%
\@ifundefined{embw}{}{}\@ifundefined{wtmp}{}{}
\makeatother
\definecolor{slotI}{RGB}{37,99,235}\definecolor{slotII}{RGB}{13,148,136}\definecolor{slotIII}{RGB}{217,119,6}

\providecommand{\embctx}{\tikz[baseline=0pt,x=1mm,y=1mm]{\useasboundingbox (0,0) rectangle (2.4,2.4); \fill[black] (1.2,1.12) circle (1.0mm);}}

\providecommand{\embna}{\tikz[baseline=0pt,x=1mm,y=1mm]{\useasboundingbox (0,0) rectangle (2.4,2.4); \draw[black,line width=0.22pt,fill=white] (1.2,1.12) circle (1.0mm);}}

\newcommand{\twolinecell}[2]{%
  \begin{tabular}[t]{@{}l@{}}#1\\#2\end{tabular}%
}

\definecolor{codegreen}{rgb}{0,0.6,0}
\definecolor{codegray}{rgb}{0.5,0.5,0.5}
\definecolor{codepurple}{rgb}{0.58,0,0.82}
\definecolor{backcolour}{rgb}{0.95,0.95,0.96}

\lstdefinestyle{ccodestyle}{
    backgroundcolor=\color{backcolour},   
	commentstyle=\color{codegreen},
	keywordstyle=\color{magenta},
	numberstyle=\tiny\color{codegray},
	stringstyle=\color{codepurple},
	basicstyle=\ttfamily\footnotesize,
	language=C,
	basicstyle=\fontsize{6}{6}\ttfamily,
	breakatwhitespace=false,         
	breaklines=true,                 
	captionpos=b,                    
	keepspaces=true,                 
    xleftmargin=10pt,
	numbers=left,                    
	numbersep=6pt,                  
	showspaces=false,                
	showstringspaces=false,
	showtabs=false,                  
	tabsize=2
}

\makeatletter
\lst@Key{highlightlines}{}{}
\makeatother

\lstnewenvironment{ccode*}[1]
    {\lstset{style=ccodestyle,#1}}
    {}

\newtcolorbox{boxI}{
    colback = lightgray!10, 
    colframe = black, 
    boxrule = 0.5pt, 
    toprule = 0.5pt, %
    arc = 2pt,
    left = 1pt,
    right = 1pt,
    bottom = 0pt,
    top = 0pt,
    before skip = \smallskipamount,
    after skip = \smallskipamount,
}

\newcounter{observcntr}

\tikzset{
  icnbase/.style={
    inner xsep=1.3pt, inner ysep=1.2pt,
    outer sep=0pt, line width=0.4pt,
    font=\bfseries\fontsize{7}{7}\selectfont,
    text height=1.3ex, text depth=0pt,
    rectangle, sharp corners
  },
  pcnrect/.style={icnbase},
  icnrect/.style={icnbase},
  icnfill/.style={fill=black, draw=black, text=white},
  icnline/.style={fill=white, draw=black, text=black},
}

\newcommand{\PLbl}[1]{{\rmfamily P}\kern0.06em{\sffamily #1}}
\newcommand{\ILbl}[1]{{\rmfamily I}\kern0.06em{\sffamily #1}}

\DeclareRobustCommand{\PB}[1]{\tikz[baseline=(n.base)]{\node[pcnrect,icnfill](n){\PLbl{#1}};}}

\DeclareRobustCommand{\IW}[1]{\tikz[baseline=(n.base)]{\node[icnrect,icnline](n){\ILbl{#1}};}}

\newcommand{\pgap}{\kern0.12em}

\DeclareRobustCommand{\vsep}{%
  \kern0.14em\raisebox{-0.15ex}{\rule{0.7pt}{1.6ex}}\kern0.14em}

\usetikzlibrary{arrows.meta}

\DeclareRobustCommand{\shortlarr}{%
  \kern0.14em%
  \raisebox{0.55ex}{%
    \begin{tikzpicture}[baseline=0pt]
      \draw[line width=0.7pt, -{Stealth[length=3.6pt, width=3.6pt]}, black]
        (0.55em,0) -- (0,0);
    \end{tikzpicture}}%
  \kern0.14em}

\newcommand{\ib}{\kern0.1em\allowbreak}

\begin{document}

\title{SoK: AI-Augmented Binary Reversing}

\author{\IEEEauthorblockN{Yujeong Kwon\textsuperscript{\dag} \quad
Yiyue Zhang\textsuperscript{\dag} \quad
Kexin Pei\textsuperscript{\ddag} \quad
Dokyung Song\textsuperscript{\S} \quad
Hyungjoon Koo\textsuperscript{\dag}}

\IEEEauthorblockA{\textsuperscript{\dag}\textit{Sungkyunkwan University} \qquad \textsuperscript{\ddag}\textit{The University of Chicago} \qquad \textsuperscript{\S}\textit{Yonsei University}}

\thanks{Corresponding author: Hyungjoon Koo (kevin.koo@skku.edu)}
}

\maketitle

\begin{abstract}

Binary reversing is fundamental to 
software understanding, vulnerability discovery, 
malware investigation, and firmware auditing.
However, it remains inherently challenging 
due to the %
\UP{lossy transformation of} semantic 
information during compilation. 
Recent advances in machine learning, 
large language models (LLMs), and
agentic AI systems have accelerated 
the adoption of AI-augmented binary reversing.
Yet, the resulting body of work 
has become increasingly 
fragmented across reversing domains, 
artifact representations, 
learning approaches, and evaluation practices.
This paper presents the first comprehensive 
systematization of knowledge on
AI-augmented binary reversing. 
We \UP{collect 246} research papers published since 2015,
and organize them into 22 binary reversing domains 
according to the inference tasks. 
We further introduce a unified taxonomy spanning 
conventional and AI-augmented reversing pipelines. 
Our taxonomy connects traditional analysis techniques, 
binary-derived artifacts, representation strategies, 
learning paradigms, and downstream inference tasks, 
while clarifying the emerging roles of 
LLMs and agentic AI systems.
By establishing a common vocabulary and 
structured framework, we \UP{offer} a holistic 
view of the field's evolution over the past decade.
Our study reveals common structures underlying
seemingly disparate approaches, highlights persistent 
technical challenges and evaluation gaps, 
and identifies promising opportunities for future research. 
Collectively, these insights clarify the current 
state of the field and provide
a foundation for the next generation of 
\UP{evidence-grounded and practically deployable}
AI-augmented binary reversing systems.

\end{abstract}

\section{Introduction}
\label{sec:intro}

As modern society increasingly relies on 
\UP{software distribution}
whose source code is unavailable 
or inaccessible, understanding the inner workings 
of executable binaries has become a critical capability.
\UP{Broadly,} binary reverse engineering 
(hereinafter referred to as {\em binary reversing}) is the \UP{analysis} of 
\UP{compiled binary artifacts to recover,
infer, understand, or characterize 
higher-level information about programs
(\eg structures, semantics, functionalities,
provenance, and security properties)}
from low-level program \UP{representations}
(\eg machine instructions, control flows, 
and data flows).
This process plays a crucial role in 
securing computing systems
by enabling vulnerability discovery~\cite{shoshitaishvili2015firmalice,zhao2024leveraging,P130_FICS2021usenix,P135_PreFuzz2022icse,P134_mtfuzz2020fse}, malware investigation~\cite{kolbitsch2010inspector,comparetti2010identifying}, 
attack comprehension~\cite{xu2014autoprobe,caballero2009dispatcher,see2023binary}, software supply-chain 
risk assessment~\cite{zhao2023one,P025_binaryai2024icse}, and patch validation~\cite{wu2025veribin,zhang2018precise,xu2023patchdiscovery}.
Beyond practical security applications, 
binary reversing is indispensable for 
software maintenance~\cite{duan2019automating}, interoperability~\cite{kraft2005toward}, 
digital forensics~\cite{lewis2018memory,maggio2021seance,cohen2015characterization,hand2012bin}, intellectual-property verification~\cite{rosenblum2011wrote,P070_caliskan2015ndss,tian2015software,duan2017identifying}, 
and legacy system migration~\cite{cifuentes1996binary,canfora2000decomposing,fujiwara2016reverse}. 

Despite its importance, binary reversing 
remains fundamentally challenging.
Compilation irreversibly discards 
high-level semantic information, 
while compiler optimizations further 
transform program structures, causing 
semantically equivalent source code 
to yield substantially different binaries 
across architectures, compilers, 
and optimization settings.
Moreover, many binary reversing
\UP{domains} require reasoning about program behavior 
and are subject to well-known 
computability limits, rendering perfect 
automated semantic recovery infeasible 
in general.
The challenge is further amplified by the scale 
and complexity of modern software, as well as 
adversarial techniques such as obfuscation, 
packing, encryption, and anti-analysis mechanisms.
As a result, binary reversing is not 
a single task but \emph{a family of 
interdependent program semantic
inference problems} 
that require combining 
evidence across 
multiple analysis modalities 
and abstraction levels.

For decades, researchers have addressed
these challenges through triage~\cite{david2017structural,kanj2026automating,mahdi2024detection,zatloukal2017malware}, 
static analysis~\cite{balakrishnan2004analyzing,lin2021function,muntean2018cfi}, 
dynamic analysis~\cite{haller2016scalable,sang2024airtaint,zhang2024hardtaint,abdelwahed2023detecting}, 
symbolic reasoning~\cite{cha2012unleashing,xu2017cryptographic,poeplau2020symbolic,tempel2025accurate}, and
program testing~\cite{lyu2019mopt,aschermann2019redqueen,pham2016model,huang2020pangolin,liu2022automated}, supported by a wide range of
automated techniques and tools~\cite{ghidra,ida,balakrishnan2005codesurfer,angr,avgerinos2011tie,slowinska2011howard,lin2010automatic}. 
Unfortunately, these approaches only partially alleviate the 
core difficulties of binary reversing. Specifically, static and symbolic analyses 
must approximate behaviors that are undecidable or infeasible to 
recover precisely, dynamic analyses and testing observe only explored 
executions, and all of these techniques operate on binaries from which 
names, types, abstractions, and design intent may have been irreversibly lost.
As a result, practical reversing still relies heavily on human experts 
to reconcile incomplete, ambiguous, and heterogeneous evidence.

This persistent gap between what automated analyses can reliably recover 
and what analysts need to understand has made machine learning (ML)
increasingly attractive 
as complementary mechanisms for augmenting binary reversing workflows.
In particular, the remarkable success of 
deep learning has 
fueled the adoption of 
\UP{AI techniques in binary reversing} 
across every stage of the 
binary-reversing pipeline, 
from low-level code recovery to high-level 
semantic understanding.
These approaches aim to learn patterns from 
binary-derived artifacts, 
automate analysis tasks, and assist 
analysts in \UP{inferring 
higher-level information about programs}.

However, the rapid adoption of ML,
LLMs and related AI techniques has produced 
a large and increasingly \emph{fragmented} 
body of research 
on AI-driven binary reversing
across a broad spectrum of domains.
Although individual studies have demonstrated 
promising results on specific tasks, 
this proliferation has outpaced our collective
understanding of the field, 
leaving its methodological
foundations, evaluation practices, 
\UP{strengths and limitations of 
an individual approach},
open challenges, and \UP{future directions} 
largely unsystematized.
This lack of synthesis has concrete consequences:
\UP{\WC{1}}~studies often reach apparently conflicting conclusions (\S\ref{sss:repr-val})
because they adopt different task formulations, datasets, 
baselines, and evaluation protocols; 
\UP{\WC{2}}~similar modeling ideas 
are repeatedly repurposed across adjacent applications (\S\ref{sss:model-val})
without a clear account of what is genuinely transferable; 
and \UP{\WC{3}}~reported progress frequently remains concentrated on 
benchmark-level improvements (\S\ref{sss:corpus-val}) rather than integration into 
real-world reversing workflows.
\UP{Consequently}, it remains difficult for researchers and 
practitioners to assess which AI techniques are robust, 
where they provide meaningful analyst assistance, and 
which barriers must be overcome for practical deployment.
To provide such a consolidated view, 
this SoK organizes the fragmented literature 
\UP{on AI-augmented binary reversing} around 
\UP{four} research questions:
\UP{
\textbf{(RQ1)} How can the fragmented literature 
be systematized as a common 
pipeline~(\Cref{fig:ai-overview})? 
\textbf{(RQ2)} How much has AI/ML augmented binary reversing 
across domains, artifacts, inference
forms, and semantic 
depth~(\S\ref{ss:rq1-landscape-maturity})?
\textbf{(RQ3)} What binary-specific validity risks 
arise from using binary corpora, 
tool-produced artifacts, representation choices, 
proxy labels, learning models, and 
evaluation metrics to support semantic 
claims~(\S\ref{ss:rq2-validity-risks})?
\textbf{(RQ4)} What open challenges and 
research directions 
are most critical for reliable 
AI-augmented reversing~(\S\ref{sec:future})?
}

Our key observation is that 
AI-augmented binary reversing
complements, rather than replaces,
conventional reversing,
and the two approaches mutually 
reinforce one another.
Simply put, conventional analysis techniques 
produce binary-derived artifacts
through a variety of reversing techniques,
subsequently serving as 
learning data for AI systems.
\UP{The outputs of these AI systems, in turn, 
provide feedback that supports further analysis.}
Viewed through this lens, 
we introduce a \emph{two-pipeline model
connected by an artifact interface}; namely, 
executable binaries are translated
into analysis artifacts, artifacts into 
AI-consumable representations, 
and representations into 
inferred semantic properties.
This artifact-centric and inference-oriented 
perspective reveals a common workflow
underlying seemingly disparate approaches 
and provides a principled basis for reasoning 
about their capabilities, limitations, and 
opportunities afforded by AI systems.

Taking this perspective, 
\UP{we investigate a corpus of 246 research papers 
published since 2015, drawn from 11 top-tier security, 
software engineering, and AI venues and supplemented 
with additional relevant studies identified 
through snowballing.}
We then organize these works into 
22 binary reversing domains
\UP{according to their primary
inference tasks.}
\UP{By providing a holistic view of the relationships 
among analysis artifacts, representation strategies, 
learning paradigms, and evaluation methodologies, our 
systematization highlights well-studied and 
underexplored domains, persistent technical challenges, 
and promising research directions.}
Furthermore, we provide an interactive
visualization 
platform\footnote{\url{https://sok-aire.vercel.app}}
(\Cref{fig:web-example} in Appendix)
to support researchers and practitioners 
in developing
reliable, \UP{evidence-grounded, and 
practically deployable}
AI-augmented binary reversing systems.

The main contributions of our paper are as follows:

\begin{itemize}[leftmargin=*]
    \item \textbf{A comprehensive paper corpus and reversing domains}
    \UP{We curate 246 papers published 
    since 2015 and organize them into 
    22 binary reversing domains by inference type, 
    providing a holistic
    view of a rapidly fragmented literature.}
    \item \textbf{An artifact-mediated taxonomy
    spanning conventional and AI pipelines:}
    \UP{We introduce a unified two-pipeline model 
    that connects conventional reversing 
    to AI-driven inference through
    binary-derived analysis artifacts.}
    \item \textbf{Binary-specific agenda and future directions:}
    \UP{We derive 12 binary-reversing-specific
    insights and 
    identify research directions
    toward end-to-end binary comprehension.}
\end{itemize}

\section{Methodology and Scope}
\label{sec:literature}
\PP{Literature Collection}
We first collected candidate papers using 
domain-specific keywords in Google 
Scholar~\cite{google_scholar} and 
screened publications from 11 top-tier 
\UP{conferences} in security (S\&P, CCS, 
USENIX Security, NDSS), software 
engineering (ICSE, FSE), and 
artificial intelligence (AAAI, IJCAI, 
NeurIPS, ICLR, ICML) published since 2015.
We chose \UP{that year} as the starting point 
because it marked the publication of 
Shin et al.'s~\cite{P005_shin2015usenix} 
function boundary detection work, 
one of the pioneering 
deep-learning-based approaches 
for identifying function boundaries 
in stripped binaries.
For each domain, we constructed
search queries by combining domain-specific
synonyms with binary-related terms.
For example,
function-name recovery queries included
alternative formulations (\eg function
name prediction, function symbol name
recovery, binary, executable, and
stripped binaries) to improve
recall despite terminological variation.
We included studies whose 
primary contribution \UP{involves applying}
ML, LLMs, or other AI techniques to 
binary reversing tasks.
We then expanded the corpus through 
backward and forward snowballing,
\UP{incorporating studies outside 
the initial venue set.}
\UP{
To avoid disadvantaging recent work, we used 
a publication-year-adjusted citation threshold of 
12 citations for 2015-2016 papers, decreasing 
by two citations every two publication years 
to 2 citations for 2025-2026 papers 
(available up to July 31, 2026).
}
This process enhances 
coverage and reduces selection bias 
and the risk of overlooking influential 
research published 
in other venues.
Our final corpus comprises 
\UP{246} research papers 
from \UP{87} venues. %

\PP{Scope}
\UP{
We include studies in which binary-derived 
artifacts are required during inference or 
evaluation and the output supports 
executable-level structure, behavior, provenance, 
security semantics, or analyst understanding.
In particular, we exclude approaches
that rely on source code at inference time,
restricting our scope to 
binary reversing tasks.
}
Finally, we limit our scope to 
native executables produced 
by compilers. 
Consequently, Android bytecode 
(\eg Java/Dalvik applications) and 
interpreted-language bytecode 
(\eg Python) fall outside the scope 
of this study.

\PP{Related Works}
Existing surveys of AI-augmented binary analysis focus on individual domains, including binary code similarity~\cite{haq2021survey}, 
information (\eg function signatures, variable types, and names) recovery~\cite{shao2022survey}, 
malware detection and classification~\cite{singh2021survey,song2025application}, 
and vulnerability detection~\cite{adhikari2025survey,shen2020survey}.
Although valuable within their respective areas, they provide no unified cross-domain perspective, 
and the field still lacks a comprehensive systematization of AI-augmented binary reversing.
\UP{
Another line of work examines 
the methodology of learning-based security research, 
identifying recurring pitfalls in experimental design 
and evaluation and measuring their prevalence 
in practice~\cite{arp2022dos,evertz2026chasing}.
While several validity concerns overlap 
with these pitfalls, 
our focus differs in how these risks manifest: 
we examine how they arise in binary reversing, 
where models operate on binary-derived artifacts,
rather than treating them as 
generic machine-learning issues.
}

\section{A Taxonomy for Binary Reversing}
\label{sec:bin-reversing-taxonomy}

\begin{figure*}[t!]
	\centering
    \includegraphics[width=0.96\textwidth]{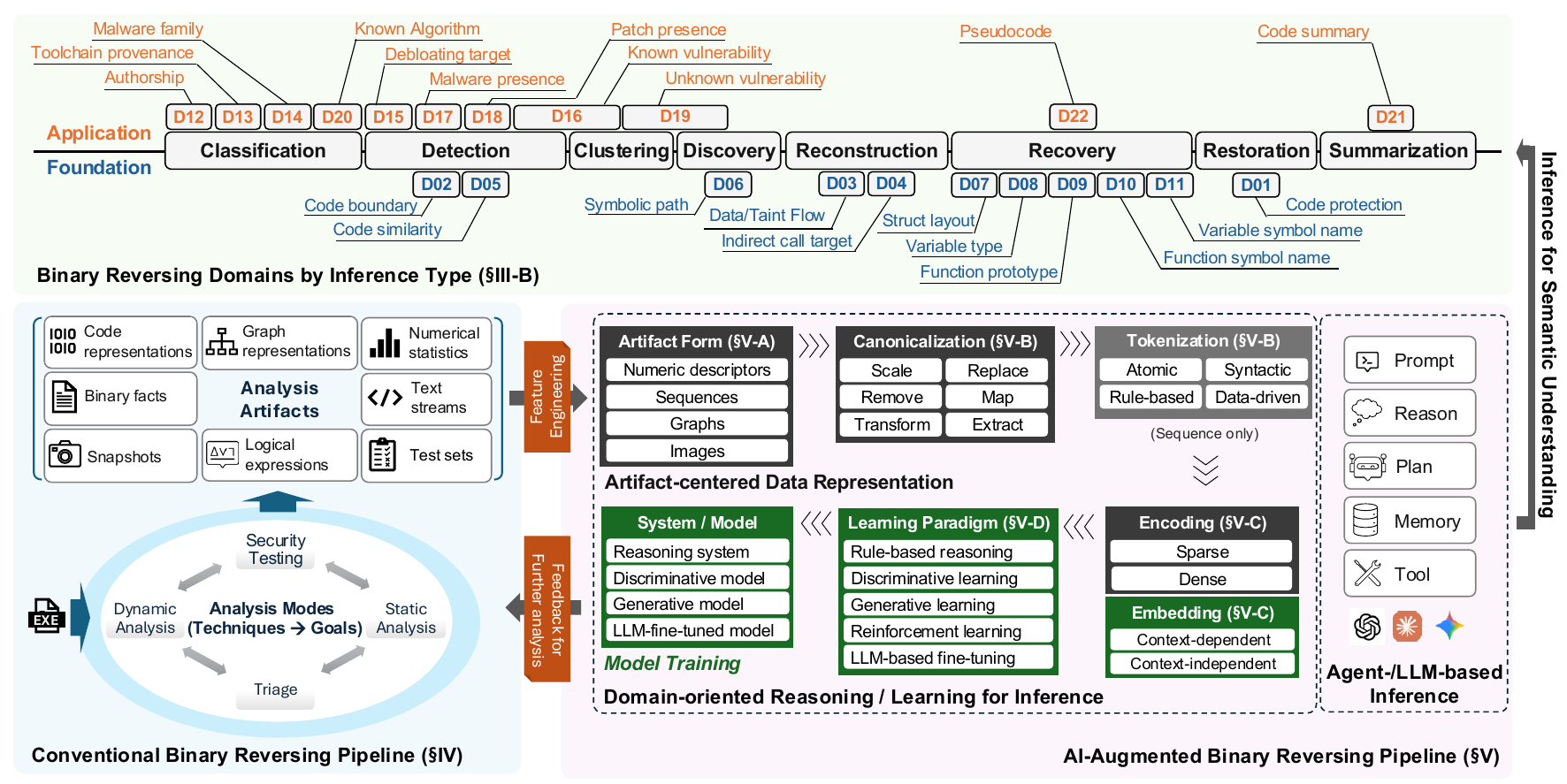}
	\caption{
    \UP{A streamlined overview of the 
    artifact-mediated pipeline that connects
    conventional binary reversing (\Cref{s:cbrp}) 
    to AI-augmented inference 
    (\Cref{sec:ai-augmented}).}
    \UP{We use this pipeline to systematize 
    fragmented AI-augmented binary reversing 
    studies across 22 domains 
    by inference type (\Cref{subsec:ai-tasks}).}
    The conventional pipeline is 
    structured around four analysis modes
    (\Cref{tbl:core-reversing-taxonomy}): 
    triage, static analysis, 
    dynamic analysis, and 
    security testing, each of 
    which is associated with distinct 
    analysis goals and 
    techniques.
    These analyses produce eight classes 
    of analysis artifacts, such as 
    code and graph representations
    (\Cref{tbl:analysis-artifact-taxonomy}), which 
    \emph{serve as the interface} 
    (\Cref{subsec:artifact-forms}) 
    to the AI-augmented pipeline. 
    Earlier approaches transform 
    them through data representation
    (\Cref{subsec:canonicalization}--\Cref{subsec:encoding-embedding})
    and model-learning processes
    (\Cref{subsec:learning-paradigms})
    to enable downstream inferences, 
    whereas recent approaches 
    exploit the prompting, reasoning, planning, 
    and \UP{tooling} capabilities 
    of LLMs and agentic AI systems. 
    \UP{Further,} the resulting inferences 
    can either guide subsequent 
    analyses or assist in 
    understanding underlying 
    code semantics.
    }
	\label{fig:ai-overview}
\end{figure*}

\subsection{Taxonomy Overview}
\label{ss:tax-overview}

\PP{Two Pipelines and Artifact Interface}
The conventional binary reversing 
pipeline differs fundamentally 
from its AI-based counterpart:
the former relies primarily 
on binary analysis techniques, 
explicit reasoning, and human expertise, 
whereas the latter introduces 
learning-based mechanisms that 
automate 
part of this 
inference process by leveraging patterns 
learned from massive datasets.
Although both pipelines share the common 
objective of recovering high-level
code semantics, 
the AI-centric paradigm 
enables more scalable analysis and 
rapid triage while reducing the dependence 
on domain expertise and manual effort.
At its core, the transition is driven by
binary-derived artifacts that 
traditional analyses and reversing tools
produce. 
These artifacts provide 
structured observations that 
can be transformed into model-consumable 
representations and embedded in 
high-dimensional feature spaces.
Such transformation naturally recasts 
a wide range of binary analysis tasks 
as inference problems 
amenable to learning-based approaches, %
including detection, classification, 
clustering, recovery, 
retrieval, and summarization.
In essence, the conventional pipeline 
describes how analysis 
artifacts are generated, 
whereas the AI-augmented 
pipeline leverages the artifacts for 
representation learning and
downstream semantic inference.
Crucially, binary-derived artifacts 
constitute the \emph{interface 
between the two pipelines}.

\PP{Taxonomy Overview}
\Cref{fig:ai-overview} 
presents our taxonomy of binary
reversing. 
We structure the conventional 
binary reversing pipeline 
around four recurring analysis modes:
triage, static analysis, dynamic analysis, 
and security testing. 
Each mode has distinct analysis objectives and 
supporting techniques and each produces %
analysis artifacts that facilitate downstream
code-semantic recovery.
These artifacts form the interface 
between the conventional and AI-augmented
pipelines. 
To consume these artifacts 
as learning data, 
the AI-augmented binary reversing pipeline
applies a non-trivial representation 
process (\ie feature engineering):
artifact selection, 
canonicalization, tokenization, 
encoding, and embedding.
Subsequent model-learning 
choices determine 
how the feature representations
support eight major inference tasks: 
classification, detection, clustering, 
discovery, reconstruction, recovery, 
restoration, and summarization.
Earlier AI-augmented approaches 
primarily learn 
from engineered representations of 
binary-derived artifacts, whereas 
recent approaches (\eg LLM- and 
agent-based AI systems) 
additionally leverage
reasoning, planning, prompting, 
and tool-use capabilities to coordinate
analysis steps across heterogeneous 
sources of evidence.
In both cases, the analysis insights 
either guide subsequent reversing
activities or assist analysts 
in understanding underlying 
code semantics.

\subsection{Reversing Domains by Inference Type}
\label{subsec:ai-tasks}

\begin{table}[t!]
\footnotesize
	\centering
	\caption{
    A taxonomy of AI-augmented binary
    reversing domains. 
    We define 22 reversing domains with
    8 inference types in 
    \Cref{subsec:ai-tasks}.
    The rightmost column reports the number of 
    surveyed papers assigned to each domain.
    }
	\resizebox{1\columnwidth}{!}{
		\renewcommand{\arraystretch}{1.0}
\setlength{\tabcolsep}{2.6pt}
\begin{tabular}{cllr}
\toprule
\textbf{DID} & \textbf{Category} & \textbf{Binary Reversing Domain by \textit{Inference Type}} & \textbf{\#} \\
\midrule
D01 & Foundation & Code protection \textit{restoration} & 0 \\
D02 & Foundation & Code boundary (instruction/function) \textit{detection} & 6 \\
D03 & Foundation & Data (taint) flow  \textit{reconstruction} & \UP{7} \\
D04 & Foundation & Indirect call target \textit{reconstruction} & 1 \\
D05 & Foundation & Code similarity \textit{detection} & \UP{37} \\
D06 & Foundation & Symbolic path \textit{discovery} & 1 \\
D07 & Foundation & Data structure/struct layout \textit{recovery} & \UP{6} \\
D08 & Foundation & Variable type \textit{recovery} & \UP{8} \\
D09 & Foundation & Function signature \textit{recovery} & \UP{4} \\
D10 & Foundation & Function symbol name \textit{recovery} & \UP{9} \\
D11 & Foundation & Variable symbol name \textit{recovery} & 6 \\
\midrule
D12 & Application & Programmer authorship \textit{classification} & \UP{3} \\
D13 & Application & Toolchain provenance \textit{classification} & \UP{8} \\
D14 & Application & Malware family \textit{classification} & \UP{37} \\
D15 & Application & Debloating target \textit{detection} & \UP{3} \\
D16 & Application & Known vulnerability \textit{detection/clustering} & \UP{10} \\
D17 & Application & Malware \textit{detection} & \UP{41} \\
D18 & Application & Patch presence \textit{detection} (binary diffing) & \UP{4} \\
D19 & Application & Unknown vulnerability \textit{discovery/clustering} & \UP{13} \\
D20 & Application & Known algorithm \textit{classification} & 2 \\
D21 & Application & Code \textit{summarization} & \UP{6} \\
D22 & Application & Pseudocode \textit{recovery} (decompilation) & \UP{12} \\
\midrule
- & Others & \makecell[l]{XAI (\UP{4}), Code representation learning (\UP{12})\\Dataset drift (\UP{5}), Obfuscated code generation (1)} & \UP{22} \\
\midrule
\multicolumn{3}{r}{\textbf{Total}} & \textbf{\UP{246}} \\
\bottomrule
\end{tabular}

	}
	\label{tbl:ai-task-taxonomy}
\end{table}

\PP{Inference Types}
We identify eight inference types in binary reversing domains:
\WC{1} \emph{classification}, which
assigns a target to one of
predefined categories;
\WC{2} \emph{clustering}, which
groups similar instances 
into emergent unlabeled clusters;
\WC{3} \emph{detection}, which
determines whether a property is present;
\WC{4} \emph{discovery}, which identifies
previously unknown candidates of interest
for subsequent validation; 
\WC{5} \emph{reconstruction}, which infers 
latent structural relations;
\WC{6} \emph{recovery}, which recovers 
source-level semantics lost \UP{in} compilation;
\WC{7} \emph{restoration}, which 
reverses intentional concealment by
code or data transformations; and 
\WC{8} \emph{summarization}, which
generates concise semantic descriptions.
Papers outside our taxonomy are grouped 
under \emph{Others}.

\PP{Domains}
\emph{Foundation domains} focus on recovering, 
reconstructing, or restoring binary-derived artifacts 
and program properties, whereas 
\emph{application domains} 
leverage such artifacts to infer higher-level semantics, 
security, provenance, or explanatory insights. 
This distinction highlights a common progression 
from artifact-centric inference to analyst-oriented 
reasoning across AI-augmented binary reversing. 
When a paper spans multiple domains, 
we assign it to the domain that 
best reflects its primary inference task.
\Cref{tbl:ai-task-taxonomy} organizes 
AI-augmented binary reversing  
according to inference type and scheme,
which yield 22 binary-reversing domains.
A small number of additional studies 
fall outside of our predefined reversing domains,
addressing explainable AI (XAI) techniques~\cite{P115_finer2023ccs,P127_lemna2018ccs,p249_d23,p250_d23},
representation learning~\cite{P033_palmtree2021ccs,P036_codeart2024fse,P037_clap2024issta,P052_hext52023ase, P116_MAIE2023usenix,P149_Nova2025iclr,p251_d24,p252_d24,p253_d24,p254_d24,p255_d24,p256_d24}, dataset drift~\cite{P111_nguyen2022aaai,P118_thirumuruganathan2024usenix,P125_cade2021usenix,p257_d25},
or code obfuscation~\cite{P001_METAMORPHASM2025aaai}.
As a final note, we deliberately retain domains 
with only a few studies to 
make underexplored areas visible.

\section{Conventional Binary Reversing}
\label{s:cbrp}

\begin{table*}[t!]
	\centering
	\caption{
    Four recurring analysis modes in conventional binary analysis: 
    triage, static analysis, dynamic analysis, and security testing~(\Cref{s:cbrp}). 
    Each analysis mode leverages
    a set of analysis techniques to achieve particular objectives,
    producing corresponding analysis artifacts.
    We curate the techniques and artifacts
    from the AI-augmented binary
    reversing literature 
    rather than seeking to provide an 
    exhaustive catalog.
    }
    \fontsize{7}{8}\selectfont
	
    \begin{tabularx}{0.99\textwidth}{@{}
>{\RaggedRight\arraybackslash}p{0.06\textwidth}
@{\hspace{1.3pt}}
>{\RaggedRight\arraybackslash}p{0.16\textwidth}
@{\hspace{13.5pt}}
>{\RaggedRight\arraybackslash}p{0.30\textwidth}
@{\hspace{16pt}}
Y@{}}
\toprule
\textbf{Mode} & \textbf{Analysis Goal} & \textbf{Representative Technique(s)} & \textbf{Analysis Artifact(s)} \\
\midrule
Triage & \makecell[tl]{Structural analysis\\Binary profiling} & \makecell[tl]{Binary header and import parsing, raw byte extraction\\String and statistical extraction} & \makecell[tl]{Header, sections, imports, exports, byte\\Strings, constants, n-grams, histogram, entropy, provenance graph
\\
chi-square statistic
} \\
\midrule
Static Analysis &
\makecell[tl]{Disassembly\\Function boundary recovery\\Control-flow analysis\\Call-site analysis\\IR lifting\\Data-flow analysis\\Pointer/alias analysis\\Static taint analysis\\Control-dependence analysis\\Data-dependence analysis\\Program-dependence analysis\\Stack-frame analysis\\Decompilation} 
&
\makecell[tl]{Linear and recursive sweep, superset disassembly\\Prologue and call-target heuristics\\Control-flow graph construction\\Direct call-site identification, indirect call resolution\\Instruction lifting, SSA conversion\\Reaching definitions, def-use linking\\Andersen's and Steensgaard's algorithm\\Taint propagation\\Post-dominator computation\\Load-store dependence linking\\Dependence-graph merge, slicing\\Stack and frame pointer tracking\\Expression and AST reconstruction} &
\makecell[tl]{Assembly, instruction format fields\\Function boundaries\\CFG, IFG\\CG, ICFG\\IR\\DFG, def-use chains, memory-access facts \\Points-to relations, alias sets\\Tainted values (register, memory, callee argument)\\CDG, PDT\\DDG\\PDG\\Stack-frame layout (variable slots, offsets)\\Pseudocode, variable data layout, AST 
} \\
\midrule
Dynamic Analysis & \makecell[tl]{Execution tracing\\Forced execution\\Behavioral monitoring\\Dynamic taint analysis\\Memory forensics} & \makecell[tl]{Dynamic binary instrumentation\\Branch forcing, emulation\\API and system-call interception\\Shadow-memory tagging\\Process-memory dumping} & \makecell[tl]{Instruction/memory-access trace, instruction address\\Memory-access/runtime-value trace, instruction address\\API/syscall/system event trace,
source-sink map, provenance graph
\\Taint trace, taint map\\Memory snapshot} \\
\midrule
Security Testing & \makecell[tl]{Symbolic execution\\Fuzzing} & \makecell[tl]{State forking, SMT solving\\Mutation and coverage feedback} & \makecell[tl]{Symbolic state/constraint/expression, pre/post-state values\\Coverage map, mutated input} \\
\bottomrule
\addlinespace[2pt]
\multicolumn{4}{@{}p{\textwidth}@{}}{\footnotesize IFG: instruction-flow graph; CFG: control-flow graph; CG: call graph; ICFG: inter-procedural control-flow graph; DFG: data-flow graph; CDG/DDG/PDG: control/data/program-dependence graph; PDT: post-dominator tree; AST: abstract syntax tree; IR: intermediate representation}
\end{tabularx}

	\label{tbl:core-reversing-taxonomy}
\end{table*}

\PP{Analysis Modes} 
We taxonomize conventional 
binary reversing into four recurring
analysis modes: 
\WC{1} \emph{triage} orients 
the investigation; 
\WC{2} \emph{static analysis} 
reconstructs program structure 
without execution; 
\WC{3} \emph{dynamic analysis} 
\UP{observes} concrete executions 
and runtime behavior (\ie observational); and
\WC{4} \emph{security testing} 
actively perturbs inputs, program states, 
execution environments, or path constraints 
to validate hypotheses (\ie interventional). 
These modes interact iteratively rather than forming a strict waterfall:
a static observation may motivate a dynamic trace; 
a trace may reveal an unpacked payload for static reanalysis; and 
a failed test may send the analyst back to revise control-flow, data-flow, 
or type hypotheses. 
\Cref{tbl:core-reversing-taxonomy} details these modes
by mapping each to its analysis %
goals, representative techniques, 
and resulting analysis artifacts.

\PP{Triage}
Triage often serves as the entry point 
of the conventional binary reversing pipeline.
It aims to characterize a binary, 
establish its analysis context, and 
prioritize subsequent 
investigation efforts.
Accordingly, triage relies on 
lightweight techniques that 
inspect structural, functional, 
provenance, and protection-related 
signals rather than  
full semantic recovery.
The information obtained 
at this stage provides 
coarse but actionable artifacts 
for \UP{revealing} the nature of 
the binary (\ie profiling), 
identifying its external 
capabilities and dependencies
(\eg libraries, network \UP{activity}), 
inferring how it may have been produced or 
protected, and determining 
which hypotheses warrant further 
investigation in other analysis modes.

\PP{Static Analysis}
Static analysis aims to reconstruct 
program structure and 
approximate semantics 
without executing the binary. 
It operates over decoded code, 
lifted representations, and abstract 
program relations to recover 
control flow, data flow, 
memory layout, type information, 
dependencies, and source-like abstractions. 
The artifacts offer broad, whole-program coverage,
including execution paths that may 
never be observed at runtime.
As a result, static-analysis artifacts 
constitute the primary structural and 
semantic foundation for subsequent 
dynamic analysis, security testing, 
and AI-augmented inference.
However, their accuracy remains 
subject to uncertainty arising from 
compiler optimizations, 
indirect control flow, aliasing, 
and obfuscation.

\PP{Dynamic Analysis}
Dynamic analysis targets 
execution-dependent behaviors 
and properties
that static analysis cannot 
fully resolve. 
Its techniques observe 
concrete executions, instrument 
runtime events, monitor interactions 
with the execution environment, and 
recover program state that 
\UP{appears} only during execution.
The resulting evidence provides 
grounded behavioral insights by capturing
what the binary actually does 
under specific inputs, states, 
and environmental conditions.
These artifacts complement 
static artifacts by validating or contextualizing
recovered semantics.
However, their evidential scope is 
bounded by execution coverage and 
the inputs, states, and environments
under which the binary is observed.

\PP{Security Testing}
Security testing is 
the intervention mode of 
the conventional reversing pipeline. 
It aims to explore program behaviors, 
generate \UP{diagnostic} inputs, 
validate hypotheses, and 
stress semantic assumptions 
through controlled perturbations.
Representative techniques 
actively steer execution through fuzzing, 
symbolic or concolic execution, 
path exploration, or differential testing.
The resulting artifacts capture 
exercised execution paths, 
generated inputs, solver-derived constraints, 
coverage information, failures, 
and validation outcomes.
These artifacts allow 
hypotheses and interpretations 
derived from static and dynamic analysis 
to be validated against observed program behavior, yielding 
evidence-backed conclusions.

\begin{table*}[t!]
	\centering
	\caption{
    Analysis artifact classes and their 
    common AI-consumable forms.
    Notably, the same artifact 
    may be represented in different forms 
    depending on the learning pipeline; 
    for instance, bytes can be encoded as numeric 
    descriptors, sequences, or images~(\Cref{subsec:artifact-forms}). 
    Representative artifacts and the 
    corresponding surveyed papers are 
    shown for each class.
    The symbols \protect\iconnum, 
    \protect\iconseq, \protect\icongraph\unskip,
    and \protect\iconimg~denote numeric 
    descriptors, sequences, graphs, 
    and images, respectively.
    }
	\resizebox{1.0\textwidth}{!}{
		\renewcommand{\arraystretch}{1}
\setlength{\tabcolsep}{2.0pt}
\scriptsize
\begin{tabularx}{\textwidth}{@{}
>{\raggedright\arraybackslash}p{0.15\textwidth}
>{\centering\arraybackslash}p{0.045\textwidth}
>{\raggedright\arraybackslash}X
@{}}
\toprule
\textbf{Artifact Class} &
\textbf{Form} &
\textbf{Examples} \\
\midrule
\multirow[t]{3}{=}{Code Representations}
& \iconnum
&
Instruction format field~\cite{P003_deepdi2022usenix};
assembly~\cite{P015_bingo2016fse, P059_xfl2023sp, P062_blens2025usenix, P070_caliskan2015ndss, P075_du_extend2023ccs, P087_transcend2017usenix, P096_Genius2016ccs, P097_discovre2016ndss, P151_deepreflect2021usenix}; 
IR~\cite{P015_bingo2016fse, P059_xfl2023sp, P062_blens2025usenix, P063_debin2018ccs, P136_AIFORE2023usenix};
pseudocode~\cite{P070_caliskan2015ndss}
\\
& \iconseq
& 
Byte~\cite{P002_xda2020ndss, P004_tady2025usenix, P005_shin2015usenix, P110_MalConv22021aaai, P113_Greedy2023usenix, P117_GreedyBlock2024ccs,P157_lmlms2026ndss}
assembly~\cite{P016_OrderMatters2020aaai, P017_codecmr2020nips, P023_binaug2024icse, P027_EBM2025NeurIPS, P031_INNEREYE2019ndss, P032_Asm2Vec2019sp, P038_ProRec2024neurips, P044_stateformer2021fse, P050_EKLAVYA2017usenix, P062_blens2025usenix, P103_MDSAE2019iclr, P119_shuai2024ndss, P128_deepbindiff2020ndss, P141_bin2summary2024fse, P144_coda2019neurips, P148_BTD2023usenix,P156_bdlfqwen32026AAAI,P157_lmlms2026ndss};
IR~\cite{P095_vulhawk2023ndss};
pseudocode~\cite{P025_binaryai2024icse, P038_ProRec2024neurips, P042_resym2024ccs, P043_typeforge2025sp, P054_symgen2025ndss, P066_dirty2022usenix, P067_VARBERT2024s&p, P068_GENNM2025ndss, P146_degpt2024ndss,P156_bdlfqwen32026AAAI,P157_lmlms2026ndss,P160_firmagent2026ndss,P161_idioms2026ndss,P163_sk2decompile2026iclr}
\\
& \icongraph
& 
Assembly~\cite{P007_funprobe2023esec/fse, P022_CIDetector2024icse, P024_HermesSim2024usenix, P028_Gemini2017ccs, P030_li2019pmlr, P041_osprey2021s&p, P093_CCFG2019ccs, P098_vulseekerpro2018esec_fse, P101_uvscan2023usenix, P104_HermeScan2024ndss, P142_misum2025fse,P158_binaligner2026ndss}; 
IR~\cite{P024_HermesSim2024usenix, P043_typeforge2025sp, P048_tygr2024usenix, P049_TRex2025usenix, P104_HermeScan2024ndss, P130_FICS2021usenix,P159_circe2026SP};
pseudocode~\cite{P043_typeforge2025sp, P051_CDA2025ccs, P142_misum2025fse,P159_circe2026SP}
\\
\midrule
\multirow[t]{2}{=}{Text Streams}
& \iconnum
& 
System event trace~\cite{P107_Wang2020aaai};
taint trace~\cite{P136_AIFORE2023usenix, P148_BTD2023usenix};
string~\cite{P063_debin2018ccs, P070_caliskan2015ndss, P081_TRANSCENDENT2022s&p, P083_MORSE2023s&p, P114_Adv-MalBayes2023aaai, P122_SCB2025ndss, P136_AIFORE2023usenix, P150_embersim2023NeurIPS,P162_maltree2026icml} \\
& \iconseq
& \makecell[tl]{
Instruction trace~\cite{P012_neudep2022esec/fse, P013_deepvsa2019usenix, P057_symlm2022ccs};
memory-access trace~\cite{P044_stateformer2021fse};
runtime-value trace~\cite{P012_neudep2022esec/fse, P044_stateformer2021fse, P057_symlm2022ccs};
API event trace~\cite{P107_Wang2020aaai, P108_zhang2020aaai, P120_srdc2025aaai,P162_maltree2026icml}; \\
system event trace~\cite{P120_srdc2025aaai, P121_ERW-Radar2025ndss,P162_maltree2026icml};
taint trace~\cite{P136_AIFORE2023usenix};
string~\cite{P017_codecmr2020nips, P104_HermeScan2024ndss}
} \\
& \icongraph
& 
Instruction trace~\cite{P093_CCFG2019ccs, P148_BTD2023usenix};
memory-access trace~\cite{P093_CCFG2019ccs, P148_BTD2023usenix};
runtime-value trace~\cite{P148_BTD2023usenix};
string~\cite{P028_Gemini2017ccs}
\\
\midrule
\multirow[t]{2}{=}{Binary Facts}
& \iconnum
& 
Symbolic state~\cite{P040_Learch2021CCS};
tainted value~\cite{P059_xfl2023sp, P062_blens2025usenix};
import~\cite{P059_xfl2023sp, P062_blens2025usenix, P070_caliskan2015ndss, P081_TRANSCENDENT2022s&p, P083_MORSE2023s&p, P084_malcl2025aaai, P114_Adv-MalBayes2023aaai, P122_SCB2025ndss, P150_embersim2023NeurIPS,P162_maltree2026icml}; 
export~\cite{P081_TRANSCENDENT2022s&p, P084_malcl2025aaai, P114_Adv-MalBayes2023aaai, P122_SCB2025ndss, P150_embersim2023NeurIPS,P162_maltree2026icml};
header~\cite{P081_TRANSCENDENT2022s&p, P083_MORSE2023s&p, P084_malcl2025aaai, P114_Adv-MalBayes2023aaai, P122_SCB2025ndss, P150_embersim2023NeurIPS,P162_maltree2026icml};
section~\cite{P081_TRANSCENDENT2022s&p, P084_malcl2025aaai, P114_Adv-MalBayes2023aaai, P122_SCB2025ndss, P150_embersim2023NeurIPS,P162_maltree2026icml};
constant~\cite{P059_xfl2023sp, P062_blens2025usenix, P063_debin2018ccs};
stack-frame layout~\cite{P063_debin2018ccs, P097_discovre2016ndss}; 
pre/post-state value~\cite{P015_bingo2016fse};
source-sink record~\cite{P008_neutaint2020sp};\newline
runtime-value state~\cite{P011_TAINTINDUCE2019ndss}
\\
& \iconseq
& 
Instruction address~\cite{P012_neudep2022esec/fse, P044_stateformer2021fse};
constant~\cite{P017_codecmr2020nips};
stack-frame layout~\cite{P042_resym2024ccs, P066_dirty2022usenix}
\\
& \icongraph
& Header~\cite{P007_funprobe2023esec/fse}; 
section~\cite{P007_funprobe2023esec/fse}; 
import~\cite{P007_funprobe2023esec/fse};
constant~\cite{P028_Gemini2017ccs};
stack-frame layout~\cite{P041_osprey2021s&p, P043_typeforge2025sp};
points-to relation~\cite{P041_osprey2021s&p, P042_resym2024ccs,P159_circe2026SP}; \newline
memory-access fact~\cite{P049_TRex2025usenix};
tainted value~\cite{P104_HermeScan2024ndss}
\\
\midrule
\multirow[t]{3}{=}{Graph Representations}
& \iconnum
& CFG~\cite{P070_caliskan2015ndss, P096_Genius2016ccs};
ICFG~\cite{P015_bingo2016fse, P057_symlm2022ccs, P059_xfl2023sp, P062_blens2025usenix, P097_discovre2016ndss, P151_deepreflect2021usenix};
CG~\cite{P059_xfl2023sp, P063_debin2018ccs, P136_AIFORE2023usenix};
AST~\cite{P070_caliskan2015ndss};
def-use chain~\cite{P063_debin2018ccs};
PDG~\cite{P063_debin2018ccs}; 
\\
& \iconseq
& 
CFG~\cite{P004_tady2025usenix}; 
ICFG~\cite{P032_Asm2Vec2019sp, P141_bin2summary2024fse,P160_firmagent2026ndss};
CG~\cite{P050_EKLAVYA2017usenix, P068_GENNM2025ndss,P161_idioms2026ndss}; 
def-use chain~\cite{P141_bin2summary2024fse};
provenance graph~\cite{P109_PROVDETECTOR2020ndss}
\\
& \icongraph
& CFG~\cite{P016_OrderMatters2020aaai, P017_codecmr2020nips, P022_CIDetector2024icse, P023_binaug2024icse, P024_HermesSim2024usenix, P028_Gemini2017ccs, P030_li2019pmlr, P041_osprey2021s&p, P048_tygr2024usenix, P093_CCFG2019ccs, P095_vulhawk2023ndss, P098_vulseekerpro2018esec_fse, P119_shuai2024ndss, P142_misum2025fse,P158_binaligner2026ndss}; 
DFG~\cite{P041_osprey2021s&p, P042_resym2024ccs, P043_typeforge2025sp, P048_tygr2024usenix, P049_TRex2025usenix, P051_CDA2025ccs, P142_misum2025fse};
CG~\cite{P042_resym2024ccs, P043_typeforge2025sp, P051_CDA2025ccs}; \newline
def-use chain~\cite{P024_HermesSim2024usenix, P041_osprey2021s&p, P042_resym2024ccs, P043_typeforge2025sp, P048_tygr2024usenix, P049_TRex2025usenix, P051_CDA2025ccs, P104_HermeScan2024ndss,P159_circe2026SP};
PDG~\cite{P038_ProRec2024neurips, P063_debin2018ccs}; DDG~\cite{P104_HermeScan2024ndss, P130_FICS2021usenix};
ICFG~\cite{P007_funprobe2023esec/fse, P101_uvscan2023usenix, P104_HermeScan2024ndss, P128_deepbindiff2020ndss,P159_circe2026SP};
IFG~\cite{P003_deepdi2022usenix};
AST~\cite{P142_misum2025fse}
\\
\midrule
\multirow[t]{3}{=}{Numerical Statistics}
& \iconnum
& 
Entropy~\cite{P081_TRANSCENDENT2022s&p, P083_MORSE2023s&p, P084_malcl2025aaai, P095_vulhawk2023ndss, P114_Adv-MalBayes2023aaai, P122_SCB2025ndss, P150_embersim2023NeurIPS,P162_maltree2026icml};
histogram~\cite{P081_TRANSCENDENT2022s&p, P114_Adv-MalBayes2023aaai, P122_SCB2025ndss, P150_embersim2023NeurIPS,P162_maltree2026icml};
chi-square statistic~\cite{P121_ERW-Radar2025ndss};
coverage map~\cite{P131_AFLFast2016ccs}
\\
& \iconseq
& Entropy~\cite{P082_gibert2018aaai};
coverage map~\cite{P132_neuzz2019s&p}
\\
\midrule
\multirow[t]{3}{=}{Snapshots}
& \iconnum 
&
Taint map~\cite{P011_TAINTINDUCE2019ndss}
\\
& \iconseq
& 
Taint map~\cite{P160_firmagent2026ndss}
\\
& \iconimg
& 
memory snapshot~\cite{P162_maltree2026icml}
\\
\midrule
Logical Expressions &
\iconnum &
Symbolic expression~\cite{P015_bingo2016fse, P148_BTD2023usenix};
symbolic constraint~\cite{P148_BTD2023usenix}
\\
\midrule
\multirow[t]{3}{=}{Test Sets}
& \iconnum 
&
Mutated input~\cite{P008_neutaint2020sp}
\\
& \iconseq 
&
Mutated input~\cite{P094_Picup2023esec/fse, P132_neuzz2019s&p, P134_mtfuzz2020fse, P135_PreFuzz2022icse,P160_firmagent2026ndss}
\\
& \iconimg
&
Mutated input~\cite{P094_Picup2023esec/fse}
\\
\bottomrule
\end{tabularx}

	}
	\label{tbl:analysis-artifact-taxonomy}
\end{table*}

\PP{Analysis Artifacts}
The aforementioned four analysis modes
produce a diverse set of analysis artifacts, which
\Cref{tbl:analysis-artifact-taxonomy} 
organizes into eight classes: code, graphs, 
numerical statistics, binary facts, 
test sets, snapshots, logical expressions and text streams. 
This artifact layer constitutes
the interface between conventional 
and AI-augmented reversing.
Notably, rather than learning directly from reversing activities,  AI models operate on representations derived from these artifacts; consequently, artifact quality, fidelity, and provenance may directly influence downstream inference.

\section{AI-Augmented Binary Reversing}
\label{sec:ai-augmented}

\noindent \Cref{tab:ai-pipeline} in Appendix
provides the per-study mapping of
artifact forms, representation steps,
and learning paradigms~\UP{across the 88 top-venue studies;} the
following subsections abstract this inventory
into the recurring stages of the
AI-augmented pipeline.
For brevity, we summarize 
representative examples from \UP{88} top-venue studies
(unless otherwise stated) for each stage of
the pipeline: artifact forms and classes 
in~\Cref{tbl:analysis-artifact-taxonomy},
canonicalization techniques in~\Cref{tab:canon-taxonomy},
tokenization strategies in~\Cref{tab:tok-taxonomy},
encoding and embedding choices 
in~\Cref{tbl:encoding-embedding-taxonomy}.

\subsection{Artifact Forms as Interface}
\label{subsec:artifact-forms}

\PP{Feature Engineering for Artifact-to-Data Conversion}
From a traditional machine-learning perspective, 
\emph{feature engineering} is the process of constructing 
model-consumable features from raw data through 
feature extraction, selection, transformation, and encoding. 
In the context of binary reversing, 
we use the term more broadly 
to encompass the representation-preparation 
stages that transform binary-derived artifacts 
into learning-ready inputs.

\PP{Artifact Forms}
The first step in the AI-augmented binary 
reversing pipeline is to 
represent
selected artifacts
(\Cref{tbl:core-reversing-taxonomy}) 
in model-consumable \emph{artifact forms}.
Such a representation choice
is of significance 
because models \UP{can hardly} %
learn from raw artifacts.
\Cref{tbl:analysis-artifact-taxonomy}
shows the eight artifact classes 
(from the conventional reversing pipeline):
code representations,
graph representations, 
numerical statistics,
binary facts, test sets, snapshots,
logical expressions and
text streams.
Accordingly, we define four artifact forms
that determine what aspects of the binary
are exposed to a model: 
\emph{numeric descriptors}, which 
encode compact aggregate information
while sacrificing ordering or
relational detail; 
\emph{sequences}, which
emphasize local order and token co-occurrence;
\emph{graphs}, which capture 
structural dependencies; and \emph{images}.
Notably, mappings between artifact
classes and forms are many-to-many.
For example, assembly and pseudocode may be
summarized as numeric descriptors, linearized
as token sequences, or transformed into
relational structures such as control-flow,
syntax, or data-flow graphs.
Similarly, a CFG may be represented through
numeric graph statistics, path- or walk-based
sequences, or its native graph topology; 
whereas bytes may be represented 
as numeric values, sequences, or images.

\subsection{Canonicalization and Tokenization}
\label{subsec:canonicalization}
\subsubsection{Canonicalization}
\label{subsec:canon}
After selecting analysis artifacts, prior work 
often applies canonicalization 
because raw artifacts are often
noisy, variable, and compiler-dependent. 
To reduce representation variability and 
improve learning robustness,
we taxonomize canonicalization 
methodologies by artifact form.

\PP{Value and Token Canonicalization}
For numeric descriptors, 
canonicalization typically consists of 
feature-value scaling (\eg min-max, Z-score,
log scaling), which normalizes 
\UP{values} to comparable ranges and 
prevents \UP{large-magnitude features} 
from dominating the learning process.
For sequences, common techniques 
include \WC{1} token abstraction, which
replaces concrete registers, addresses, constants, 
function or jump targets, strings, or 
identifiers with canonical placeholders 
(\eg \cc{eax} $\rightarrow$ \cc{REG},
\cc{0x401000} $\rightarrow$ \cc{ADDR},
\cc{printf} $\rightarrow$ \cc{FUNC},
\cc{hello} $\rightarrow$ \cc{STR}), and
\WC{2} token filtering, which
removes task-irrelevant or redundant 
sequence elements.
Such operations normalize low-level 
representations by abstracting semantically 
equivalent elements and 
\UP{removing} non-essential components.

\PP{Structural and Visual Canonicalization}
For graphs, canonicalization is 
more diverse and may involve node and 
edge attribute mapping, 
structural transformation (\eg reshaping 
graph topology), 
and subgraph extraction.
Such techniques standardize graph semantics, 
enrich or simplify graph structure, and 
isolate task-relevant regions 
while preserving essential program relationships.
For images, canonicalization 
primarily consists of image-size scaling and 
pixel-value scaling, which normalize 
spatial dimensions and intensity 
ranges to produce consistent visual inputs.
Collectively, these procedures mitigate 
representation heterogeneity and incidental 
variation caused by compilation and artifact diversity, 
allowing learning algorithms to focus on 
task-relevant semantic patterns rather than 
superficial differences in artifact construction.

\begin{table}[t]
\centering
\caption{
Canonicalization techniques 
by AI-consumable artifact form. 
Canonicalization determines 
which surface  variations in an artifact 
are preserved, scaled, replaced,
removed, mapped, transformed, or 
extracted prior to learning~(\Cref{subsec:canon}). 
This table summarizes 
canonicalization techniques applied
to numeric descriptor, sequence, graph, 
and image forms.
}
\label{tab:canon-taxonomy}
\scriptsize
\setlength{\tabcolsep}{3pt}
\renewcommand{\arraystretch}{1.0}
\begin{tabularx}{\columnwidth}{@{}l >{\raggedright\arraybackslash}X@{}}
\toprule
\textbf{Form} & \textbf{Canonicalization Techniques}\\
\midrule
\iconnum
&
Feature value scaling~\cite{P008_neutaint2020sp, P059_xfl2023sp, P062_blens2025usenix, P081_TRANSCENDENT2022s&p, P083_MORSE2023s&p, P084_malcl2025aaai, P095_vulhawk2023ndss, P097_discovre2016ndss, P121_ERW-Radar2025ndss, P122_SCB2025ndss, P151_deepreflect2021usenix};\newline
feature value removal~\cite{P003_deepdi2022usenix}
\\
\midrule
\iconseq
& 
Byte value scaling~\cite{P012_neudep2022esec/fse, P013_deepvsa2019usenix, P134_mtfuzz2020fse, P135_PreFuzz2022icse};\newline
abstraction by replacement~\cite{P023_binaug2024icse, P027_EBM2025NeurIPS, P031_INNEREYE2019ndss, P032_Asm2Vec2019sp, P042_resym2024ccs, P044_stateformer2021fse, P054_symgen2025ndss, P057_symlm2022ccs, P062_blens2025usenix, P066_dirty2022usenix, P067_VARBERT2024s&p} \cite{P068_GENNM2025ndss, P095_vulhawk2023ndss, P109_PROVDETECTOR2020ndss, P119_shuai2024ndss, P120_srdc2025aaai, P128_deepbindiff2020ndss, P141_bin2summary2024fse, P144_coda2019neurips,P156_bdlfqwen32026AAAI,P161_idioms2026ndss}; \newline
filtering by removal~\cite{P012_neudep2022esec/fse, P095_vulhawk2023ndss, P103_MDSAE2019iclr, P108_zhang2020aaai, P109_PROVDETECTOR2020ndss, P120_srdc2025aaai, P121_ERW-Radar2025ndss,P162_maltree2026icml}\\
\midrule
\icongraph
& 
Structural transformation~\cite{P007_funprobe2023esec/fse, P024_HermesSim2024usenix, P041_osprey2021s&p, P042_resym2024ccs, P043_typeforge2025sp, P048_tygr2024usenix, P049_TRex2025usenix, P051_CDA2025ccs, P093_CCFG2019ccs} \cite{P101_uvscan2023usenix, P104_HermeScan2024ndss, P128_deepbindiff2020ndss, P130_FICS2021usenix, P142_misum2025fse, P148_BTD2023usenix};\newline
node attribute mapping~\cite{P003_deepdi2022usenix, P024_HermesSim2024usenix, P028_Gemini2017ccs, P030_li2019pmlr, P043_typeforge2025sp, P063_debin2018ccs, P095_vulhawk2023ndss, P098_vulseekerpro2018esec_fse, P119_shuai2024ndss, P142_misum2025fse,P158_binaligner2026ndss,P159_circe2026SP}; \newline
edge attribute mapping~\cite{P003_deepdi2022usenix, P024_HermesSim2024usenix, P028_Gemini2017ccs, P030_li2019pmlr, P043_typeforge2025sp, P063_debin2018ccs, P098_vulseekerpro2018esec_fse, P142_misum2025fse,P158_binaligner2026ndss,P159_circe2026SP}; \newline 
subgraph extraction~\cite{P130_FICS2021usenix}
\\
\midrule
\iconimg
& 
Image size scaling~\cite{P094_Picup2023esec/fse,P162_maltree2026icml}; pixel value scaling~\cite{P094_Picup2023esec/fse,P162_maltree2026icml}\\
\bottomrule
\end{tabularx}

\end{table}

\begin{table}[t]
\centering
\caption{
Tokenization taxonomy for 
sequence-based artifacts. 
Tokenization defines how a sequence 
is partitioned into model-consumable 
units prior to encoding~(\Cref{subsec:tokenization}). 
We classify tokenization strategies 
by split basis: 
atomic units, syntactic units,
rule-based decomposition, 
and data-driven segmentation.
This table summarizes 
popular tokenization techniques.
}
\label{tab:tok-taxonomy}
\scriptsize
\begin{tabularx}{\columnwidth}{@{}l >{\raggedright\arraybackslash}X@{}}
\toprule
\textbf{Split Basis} & \textbf{Token Units / Tokenization Strategies}\\
\midrule
Atomic units
& Byte~\cite{P002_xda2020ndss, P004_tady2025usenix, P005_shin2015usenix, P012_neudep2022esec/fse, P013_deepvsa2019usenix, P110_MalConv22021aaai, P113_Greedy2023usenix, P117_GreedyBlock2024ccs, P132_neuzz2019s&p, P134_mtfuzz2020fse, P135_PreFuzz2022icse};
character~\cite{P017_codecmr2020nips}
\\
\midrule
Syntactic units
& Instruction/IR-level (\eg opcode, operands) ~\cite{P012_neudep2022esec/fse, P032_Asm2Vec2019sp,P095_vulhawk2023ndss} \cite{P128_deepbindiff2020ndss, P141_bin2summary2024fse,P044_stateformer2021fse}; 
whole instruction~\cite{P023_binaug2024icse, P031_INNEREYE2019ndss, P038_ProRec2024neurips, P050_EKLAVYA2017usenix, P057_symlm2022ccs, P103_MDSAE2019iclr, P144_coda2019neurips}; 
trace-level (\eg events, offsets)~\cite{P107_Wang2020aaai, 
P108_zhang2020aaai, P121_ERW-Radar2025ndss, P136_AIFORE2023usenix}
\\
\midrule
\twolinecell{Rule-based}{decomposition}
& 
Custom rule 
(\eg layout, entropy)~\cite{P109_PROVDETECTOR2020ndss,
P066_dirty2022usenix, P082_gibert2018aaai,P094_Picup2023esec/fse}
\\
\midrule
\twolinecell{Data-driven}{segmentation}
& Subword (Byte Pair Encoding)~\cite{P016_OrderMatters2020aaai, P062_blens2025usenix, P066_dirty2022usenix, P067_VARBERT2024s&p, P119_shuai2024ndss, P148_BTD2023usenix,P156_bdlfqwen32026AAAI,P157_lmlms2026ndss,P162_maltree2026icml}; 
\newline
subword (Pretrained-LLM)~\cite{P025_binaryai2024icse, P027_EBM2025NeurIPS, P038_ProRec2024neurips, P042_resym2024ccs, P043_typeforge2025sp, P054_symgen2025ndss, P068_GENNM2025ndss} \cite{P104_HermeScan2024ndss, P120_srdc2025aaai, P146_degpt2024ndss,P160_firmagent2026ndss,P161_idioms2026ndss,P163_sk2decompile2026iclr}\\
\bottomrule
\end{tabularx}

\setlength{\tabcolsep}{3pt}
\renewcommand{\arraystretch}{1.0}
\end{table}

\subsubsection{Tokenization}
\label{subsec:tokenization}
Tokenization is primarily applied to 
sequential forms, where raw sequences 
must be partitioned into discrete 
units before learning.
Tokenization determines the basic units 
from which models learn and therefore 
plays a critical role in shaping 
(subsequent) representations.
We taxonomize prior work 
according to four tokenization 
strategies: smallest observable symbols, 
language-defined symbols, 
manually decomposed symbols, \UP{and} statistically learned symbols.

\PP{Atomic and Syntactic Units}
First, \emph{atomic units}
treats the smallest observable symbols
as tokens, such as individual bytes
or characters.
These approaches require minimal
domain knowledge and preserve
the original representation at
the finest granularity.
Second, \emph{syntactic units}
leverages the syntactic structure of code
and intermediate representations.
Representative examples include
treating entire instructions as tokens,
separating opcodes and operands
into individual elements, or
using high-level-language lexer tokens.

\PP{Rule-Based and Data-Driven Tokenization}
\emph{Rule-based decomposition}
further partitions existing tokens
according to manually specified rules.
Examples include splitting compound
identifiers at naming-convention
(\ie camel- or snake-case) boundaries
(\eg \cc{GetProcAddress}
$\rightarrow$ \cc{Get}, \cc{Proc}, \cc{Address})
or decomposing complex operands and
expressions into finer-grained subtokens
(\eg \cc{mov eax, [ebx+4]} $\rightarrow$
\cc{mov}, \cc{eax}, \cc{[}, \cc{ebx},
\cc{+}, \cc{4}, \cc{]}).
Finally, \emph{data-driven segmentation}
learns token boundaries directly
from large corpora.
Representative techniques include
byte-pair encoding (BPE),
unigram language models, and
tokenizers inherited from pretrained LLMs,
which construct subword vocabularies
based on statistical regularities
in the training data.
Collectively, these strategies represent
different tradeoffs among granularity,
vocabulary size, semantic expressiveness,
generalization ability, and reliance
on domain-specific knowledge.

\subsection{Encoding and Embedding}
\label{subsec:encoding-embedding}

\begin{table*}[t!]
	\centering
	\caption{
    Encoding and embedding 
    choices~(\Cref{subsec:encoding-embedding})
    by model-consumable artifact form.
    Encoding is a fixed, non-learned conversion 
    into model-readable values and
    is categorized (by density)
    as either sparse or dense.
    Embedding is a learned representation 
    and is categorized (by context
    dependency) as
    either context-independent or context-dependent.
    This table summarizes various encoding and embedding strategies.
    }
	\resizebox{0.99\textwidth}{!}{
		\begin{tabular}{@{}
  c
  l
  >{\raggedright\arraybackslash}p{0.4\textwidth}
  l
  >{\raggedright\arraybackslash}p{0.4\textwidth}
  @{}}
\toprule
\textbf{Form} &
\makecell[l]{\textbf{Encoding}\\\textbf{Density}} &
\textbf{Encoding Examples} &
\makecell[l]{\textbf{Embedding}\\\textbf{Method}} &
\textbf{Embedding Examples} \\
\midrule
\multirow[t]{4}{*}{\iconnum} &
Sparse &
One-hot~\cite{P063_debin2018ccs}; 
multi-hot~\cite{P075_du_extend2023ccs}
&
\makecell[tl]{Context-\\independent} &
ResNet~\cite{P095_vulhawk2023ndss};
lookup layer~\cite{P057_symlm2022ccs,P162_maltree2026icml};
LDP~\cite{P059_xfl2023sp, P062_blens2025usenix};
BoostNE~\cite{P059_xfl2023sp, P062_blens2025usenix};
Autoencoder~\cite{P059_xfl2023sp, P062_blens2025usenix, P151_deepreflect2021usenix}
\\
\cmidrule{2-5}
&
\multirow[t]{3}{*}{Dense} &
Count vectors~\cite{P070_caliskan2015ndss, P083_MORSE2023s&p, P087_transcend2017usenix, P097_discovre2016ndss, P136_AIFORE2023usenix};
VLAD~\cite{P096_Genius2016ccs};
TF-IDF~\cite{P070_caliskan2015ndss};
frequency vectors~\cite{P070_caliskan2015ndss, P131_AFLFast2016ccs}
&
\makecell[tl]{Context-\\dependent} &
RNN~\cite{P003_deepdi2022usenix};
CNN~\cite{P136_AIFORE2023usenix}
\\
\midrule
\multirow[t]{2}{*}{\vspace{0pt}\iconseq} &
\makecell[tl]{\vspace{0pt}Sparse} &
One-hot~\cite{P002_xda2020ndss, P005_shin2015usenix, P013_deepvsa2019usenix, P103_MDSAE2019iclr,P156_bdlfqwen32026AAAI,P157_lmlms2026ndss,P162_maltree2026icml,P160_firmagent2026ndss,P161_idioms2026ndss,P163_sk2decompile2026iclr}
&
\makecell[tl]{Context-\\independent} &
PV-DM~\cite{P032_Asm2Vec2019sp, P109_PROVDETECTOR2020ndss};
lookup layer~\cite{P110_MalConv22021aaai, P113_Greedy2023usenix, P117_GreedyBlock2024ccs,P162_maltree2026icml};
MLP~\cite{P134_mtfuzz2020fse, P135_PreFuzz2022icse};
CBOW~\cite{P128_deepbindiff2020ndss};
CNN~\cite{P012_neudep2022esec/fse};
TextCNN~\cite{P094_Picup2023esec/fse};
Word2Vec~\cite{P050_EKLAVYA2017usenix} 
\\
\cmidrule{2-5}
&
\multirow[t]{7}{*}{Dense} &
\makecell[tl]{\vspace{0pt}
Time-series vectors~\cite{P082_gibert2018aaai}; 
TF-IDF~\cite{P128_deepbindiff2020ndss}
} &
\makecell[tl]{\vspace{0pt}Context-\\dependent} &
\makecell[tl]{
Transformer-based~\cite{P004_tady2025usenix, P012_neudep2022esec/fse, P025_binaryai2024icse, P038_ProRec2024neurips, P042_resym2024ccs, P044_stateformer2021fse, P054_symgen2025ndss,P057_symlm2022ccs, P066_dirty2022usenix}\\ \cite{P068_GENNM2025ndss,P120_srdc2025aaai, P121_ERW-Radar2025ndss, P141_bin2summary2024fse,P156_bdlfqwen32026AAAI,P160_firmagent2026ndss,P161_idioms2026ndss,P163_sk2decompile2026iclr}; 
Hierarchical BiLSTM~\cite{P013_deepvsa2019usenix}\\ \cite{P017_codecmr2020nips};
BERT-based~\cite{P002_xda2020ndss, P016_OrderMatters2020aaai, P038_ProRec2024neurips, P062_blens2025usenix, P067_VARBERT2024s&p, P095_vulhawk2023ndss, P104_HermeScan2024ndss};
Mamba~\cite{P157_lmlms2026ndss};
\\ \cite{P119_shuai2024ndss, P141_bin2summary2024fse,P023_binaug2024icse}; 
Hierarchical-LSTM~\cite{P017_codecmr2020nips}; 
Tree-LSTM~\cite{P144_coda2019neurips}; \\
Attention-based LSTM~\cite{P107_Wang2020aaai, P148_BTD2023usenix};
Integer-LSTM~\cite{P017_codecmr2020nips}; \\
LLM2Vec~\cite{P027_EBM2025NeurIPS};
Word2Vec~\cite{P031_INNEREYE2019ndss, P141_bin2summary2024fse};
Gated-CNN~\cite{P108_zhang2020aaai};\\
GRU~\cite{P050_EKLAVYA2017usenix, P141_bin2summary2024fse}; 
RNN-VAE~\cite{P103_MDSAE2019iclr}; 
RNN~\cite{P004_tady2025usenix};
BiRNN~\cite{P005_shin2015usenix}
} 
\\
\midrule
\multirow[t]{7}{*}{\icongraph} &
\multirow[t]{2}{*}{Sparse} &
Node attribute ID~\cite{P024_HermesSim2024usenix, P063_debin2018ccs};
\newline
edge attribute ID~\cite{P024_HermesSim2024usenix,P063_debin2018ccs, P093_CCFG2019ccs}
&
\multirow{2}{*}{\makecell[l]{Context-\\independent}} &
Bag-of-Nodes~\cite{P130_FICS2021usenix}
\\
\cmidrule{2-5}
&
\multirow[t]{4}{*}{Dense} &
Adjacency matrix~\cite{P016_OrderMatters2020aaai, P038_ProRec2024neurips, P095_vulhawk2023ndss, P098_vulseekerpro2018esec_fse, P119_shuai2024ndss, P128_deepbindiff2020ndss, P130_FICS2021usenix,P142_misum2025fse}; \newline
node feature mapping~\cite{P003_deepdi2022usenix, P017_codecmr2020nips, P022_CIDetector2024icse, P023_binaug2024icse,P028_Gemini2017ccs, P030_li2019pmlr,P048_tygr2024usenix, P095_vulhawk2023ndss} \cite{ P098_vulseekerpro2018esec_fse, P128_deepbindiff2020ndss, P130_FICS2021usenix, P142_misum2025fse};
\newline
edge feature mapping~\cite{P007_funprobe2023esec/fse, P041_osprey2021s&p,P051_CDA2025ccs,P101_uvscan2023usenix,P048_tygr2024usenix};
&
\makecell[tl]{\vspace{0pt}Context-\\dependent} &
GNN~\cite{P022_CIDetector2024icse, P030_li2019pmlr}; 
GIN~\cite{P119_shuai2024ndss};
GCN~\cite{P095_vulhawk2023ndss};
RGCN~\cite{P003_deepdi2022usenix, P048_tygr2024usenix};
Bidirectional GGNN~\cite{P024_HermesSim2024usenix};
GGNN~\cite{P017_codecmr2020nips};
DeepWalk~\cite{P128_deepbindiff2020ndss};
Graph2Vec~\cite{P130_FICS2021usenix};
BERT-based~\cite{P038_ProRec2024neurips};
HMA-HMP~\cite{P142_misum2025fse};
Structure2Vec~\cite{P023_binaug2024icse, P028_Gemini2017ccs, P098_vulseekerpro2018esec_fse,P158_binaligner2026ndss};
ResNet~\cite{P016_OrderMatters2020aaai}
\\
\midrule
\makecell[tl]{\vspace{0pt}\iconimg} &
\makecell[tl]{\vspace{0pt}Dense} &
\makecell[tl]{\vspace{0pt}
Grayscale tensor~\cite{P094_Picup2023esec/fse}
}
&
\makecell[tl]{\vspace{0pt}Context-\\dependent} &
\makecell[tl]{\vspace{0pt}
CNN~\cite{P094_Picup2023esec/fse}; ResNet~\cite{P162_maltree2026icml}
}
\\
\bottomrule
\addlinespace[2pt]
\multicolumn{5}{@{}p{1.1\textwidth}@{}}
{
BiLSTM: bidirectional long short-term memory;
~BiRNN: bidirectional recurrent neural network;
~BoostNE: boost network embedding;
~CBOW: continuous bag-of-words;
~CNN: convolutional neural network;
~GCN: graph convolutional network;
~GGNN: gated graph neural network;
~GIN: graph isomorphism network;
~GNN: graph neural network;
~GRU: gated recurrent unit;
~HMA-HMP: heterogeneous mutual attention-heterogeneous message passing;
~ID: identifier;
~LDP: local degree profile;
~LSTM: long short-term memory;
~MLP: multilayer perceptron;
~PV-DM: paragraph vector-distributed memory;
~RGCN: relational graph convolutional network;
~ResNet: residual neural network;
~RNN: recurrent neural network;
~RNN-VAE: RNN-based variational autoencoder;
~TextCNN: text convolutional neural network;
~TF-IDF: term frequency-inverse document frequency;
~VLAD: vector of locally aggregated descriptors
}
\end{tabular}

	}
	\label{tbl:encoding-embedding-taxonomy}
\end{table*}

\PP{Encoding}
In our pipeline, \emph{encoding}
defines the fixed, non-learned 
interface that transforms
model-consumable forms
into concrete values that can be stored,
batched, compared, or supplied to a model.
We classify encodings by density:
\WC{1} \emph{sparse} encodings 
preserve identity or membership 
through one-hot, multi-hot,
token-, node-, and edge-ID, or adjacency
indicators, whereas 
\WC{2} \emph{dense} encodings 
represent artifacts as compact 
numerical representations, 
including statistical vectors, graph matrices,
time-series vectors, and pixel tensors.
The choice depends on artifact form,
as summarized 
in~\Cref{tbl:encoding-embedding-taxonomy}.
Numeric descriptors are commonly encoded as
one-hot or multi-hot categorical features,
count or frequency vectors, 
or TF-IDF representations;
sequences use token IDs, one-hot tokens,
time-series vectors, or TF-IDF summaries;
graphs use node and edge identifiers,
feature mappings, and adjacency matrices; 
and images are encoded as dense pixel tensors.

\PP{Embedding}
\emph{Embedding} denotes learned
representations derived from encoded
artifact forms during pretraining,
fine-tuning, task training, or adaptation.
Where encoding defines the numerical
interface, embedding determines how the
model organizes binary-derived evidence within
a latent space for downstream reasoning,
classification, retrieval, generation, or
policy learning.
As shown in
\Cref{tbl:encoding-embedding-taxonomy},
we classify embeddings by their use of neighboring context:
\WC{1} \emph{context-independent}
embeddings represent a unit with little or 
no conditioning on its
surrounding artifact context, as in lookup
layers, word2vec/CBOW-style token
embeddings, PV-DM representations, local
descriptor embeddings, and bag-of-node
graph summaries.
In contrast, \WC{2} \emph{context-dependent} 
embeddings update the representation 
using neighboring tokens,
control- or data-flow neighbors, execution
context, or spatial neighborhoods.
Examples include Transformer- and
BERT-based sequence models\UP{,} RNN/LSTM/GRU sequence encoders,
graph neural or graph-embedding models\UP{,} 
and convolutional models operating on image 
or tensor representations.
Notably, the embedding stage
produces learned abstractions that
directly condition the learning paradigms 
(\Cref{subsec:learning-paradigms}).

\subsection{Learning Paradigms}
\label{subsec:learning-paradigms}

\PP{Rule-Based Reasoning}
We define a \emph{rule-based reasoning} 
paradigm when inference over analysis artifacts
is symbolic, probabilistic, or deductive rather
than learned directly from data (\eg 
end-to-end neural training).
Early ML-driven binary reversing
systems often rely on rule-based,
solver-based, or probabilistic techniques
when domains exhibit strong
program-analysis invariants, such as calling
conventions, symbolic equivalence, 
taint-propagation rules, 
type constraints, or consistency
constraints across recovered entities.
\UP{For instance, such systems apply symbolic execution and SMT solving for cross-architecture binary code search~\cite{P015_bingo2016fse}, combine classical prediction with CRF-based structured inference for debug-information recovery~\cite{P063_debin2018ccs}, learn boolean taint-propagation rules from execution observations~\cite{P011_TAINTINDUCE2019ndss}, and perform CRF-based structural inference over function symbols~\cite{P053_punstrip2020acsac}.
}
Later systems~\cite{P041_osprey2021s&p,
P051_CDA2025ccs} continue to employ 
this paradigm for structure, type, 
and signature recovery.

\PP{Discriminative Learning}
\emph{Discriminative learning} maps artifact
representations to bounded outputs 
such as labels, scores, rankings, 
pairwise matches, or cluster assignments.
This is a \UP{dominant} paradigm as in
\Cref{tbl:ai-task-taxonomy}
when targets are verifiable.
For example, code boundary detection
classifies whether bytes or instructions
correspond to valid code or function
entry points~\cite{P002_xda2020ndss,
P004_tady2025usenix,P005_shin2015usenix};
code similarity detection
learns embeddings or matching functions
over functions, basic blocks, traces,
or graphs~\cite{P028_Gemini2017ccs,
P029_safe2019springer,
P019_jtrans2022issta,P021_TREX2022tse,P156_bdlfqwen32026AAAI};
and malware, vulnerability, and provenance
domains predict class labels,
maliciousness, vulnerability relevance,
or patch presence~\cite{P083_MORSE2023s&p,
P110_MalConv22021aaai,P095_vulhawk2023ndss,
P102_wu2023IEEE_Access,P157_lmlms2026ndss}.
Note that we include 
metric learning and contrastive learning 
because their learned representations 
ultimately support decision-oriented retrieval 
or matching.

\PP{Generative Learning}
\emph{Generative learning} trains a model to
produce artifact-like or analyst-facing
outputs (\eg names, summaries, 
source-like code) rather than selecting 
from a fixed output space.
In our taxonomy, this paradigm appears
frequently in domains where 
the desirable result is open-ended or 
textual, including function
and variable name recovery~\cite{P065_dire2019ase,P066_dirty2022usenix,P068_GENNM2025ndss}, 
code summarization~\cite{P140_BinT52023saner,
P141_bin2summary2024fse,P142_misum2025fse}, 
and decompilation-oriented pseudocode generation
~\cite{P144_coda2019neurips,P143_katz2018saner}.

\PP{Reinforcement Learning}
\emph{Reinforcement learning} applies when 
a reversing task is better formulated as
sequential decision making under feedback
than as direct label prediction.
The learned object is a policy or value
function that chooses actions, such as which
symbolic state to explore, which features to
retain, when to terminate dynamic analysis, or
which fuzzing inputs to prioritize.
For example, malware-analysis systems use actor-critic learning to decide emulation control during dynamic analysis~\cite{P107_Wang2020aaai} and deep Q-learning to select discriminative features~\cite{P124_dqfsa2019IEEE_Access}.
Coverage-guided fuzzing 
systems~\cite{P131_AFLFast2016ccs,
P133_V-Fuzz2020tcyb,P134_mtfuzz2020fse,
P136_AIFORE2023usenix} further
\UP{show} the role of reward signals,
including coverage growth, crashes, and
time-to-bug.

\PP{LLM-based Fine-tuning}
\emph{LLM-based fine-tuning} is distinguished from
general generative learning because it
starts from a pretrained language model and
updates its parameters on binary-derived artifacts.
This paradigm leverages broad code and language
knowledge acquired during pretraining while
specializing the model to
assembly, IR, decompiler output, or
domain-specific reversing tasks.
Examples include staged
adaptation for function-name recovery
~\cite{P054_symgen2025ndss}, 
variable name generation 
framework~\cite{P068_GENNM2025ndss},
fine-tuning on assembly-source 
and decompiler-output 
pairs~\cite{P147_llm4decompile2024emnlp},
and %
decoder-only foundation model for
assembly-code representation and 
generation~\cite{P149_Nova2025iclr}.
Unlike prompting-only systems, the resulting
behavior depends heavily on the adaptation objective, parameter-update 
strategy, and the quality of the fine-tuning corpus.

\PP{LLM Prompting}
\emph{LLM prompting} uses a pretrained LLM as a
fixed inference engine; no model parameters
are updated for the target reversing task.
Instead, binary-derived artifacts are
textualized into prompts, potentially augmented
with examples, chain-of-thought instructions,
tool outputs, retrieval context, or
candidate lists.
This paradigm is important because recent 
\UP{LLM-based and agentic systems} can perform
classification, recovery, explanation,
candidate selection, and hypothesis
generation over heterogeneous artifacts
without requiring task-specific training.
Representative examples include 
prompt-driven workflow for taint-oriented
vulnerability analysis~\cite{P010_LATTE2025TOSEM,P160_firmagent2026ndss}, 
LLM-based readability selection for
recovered types and 
structures~\cite{P043_typeforge2025sp,P159_circe2026SP}, and
LLM-driven refinement of decompiler output
for pseudocode recovery~\cite{P146_degpt2024ndss}.
However, the reproducibility of LLM prompting
depends heavily on prompt
templates, context construction, model
version, decoding settings, and the quality
of upstream analysis artifacts.

\subsection{Evaluation Practices}
\label{sec:assessment}
\begin{figure}[t!]
\centering
\includegraphics[width=\columnwidth]{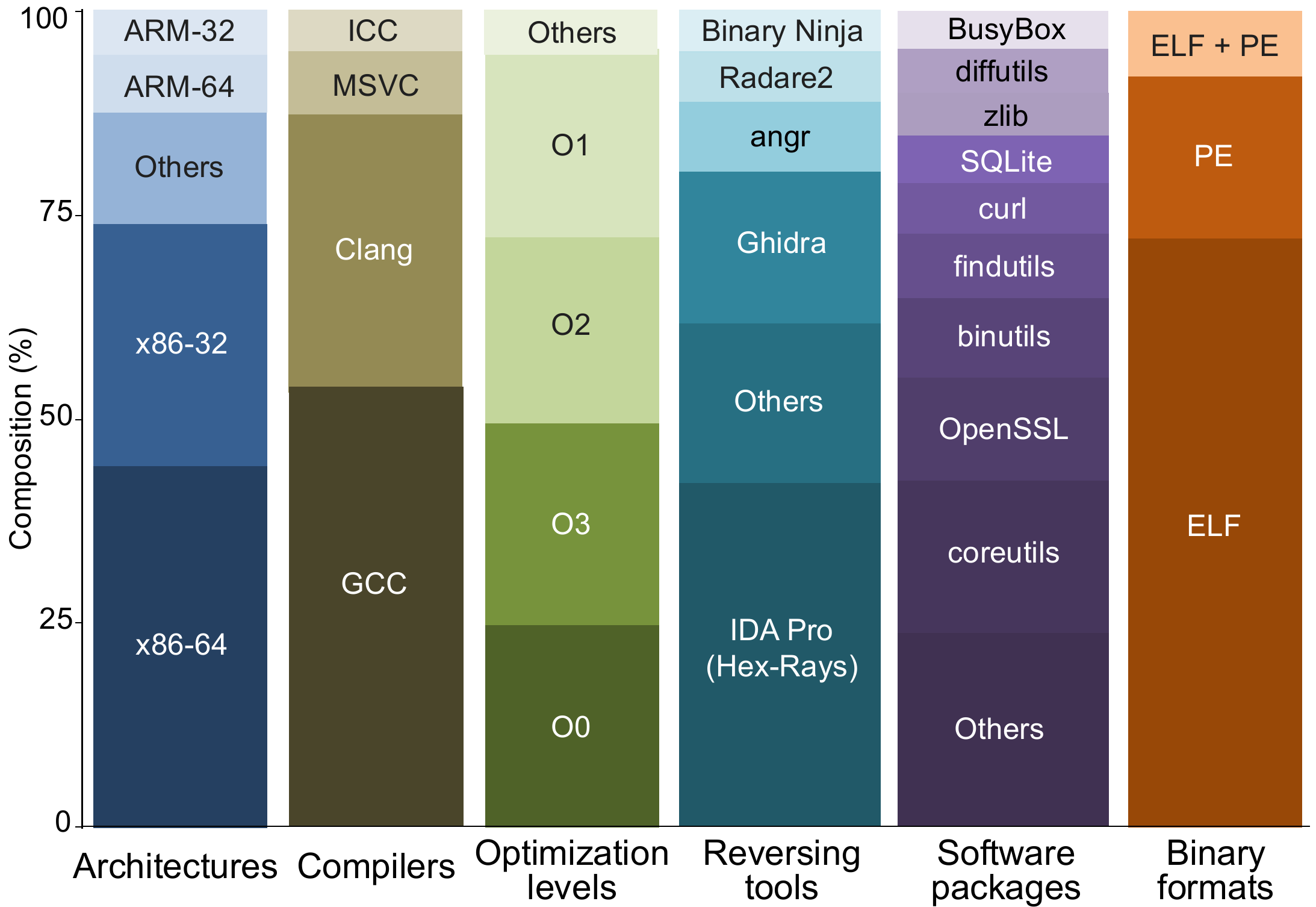}
\caption{
Approximate composition of the binary corpus 
in our paper collection, \UP{illustrating}
the most commonly 
reported architectures, compilers, optimization levels, 
reversing tools, software packages, and 
binary formats (top-k, $3 \leq k \leq 10$). 
Percentages are calculated from all reported 
occurrences rather than unique entities; 
a single option (\eg coreutils) 
contributes multiple counts.
Note that each bar is independently
normalized to 100\% and may not reflect 
the full paper corpus. 
}
\label{fig:plot-statistic}
\end{figure}

\PP{Evaluation Setup}
\Cref{fig:plot-statistic} summarizes the
reported characteristics of evaluation setups
along six
independently normalized categories: target
architecture, compiler, optimization level,
reversing tool, software package, and binary format.
Percentages are computed over reported
occurrences rather than unique papers,
binaries, or packages.
For target architectures, x86-family binaries
(x86-32 and x86-64) account for approximately
\UP{74\%} of architecture occurrences,
with ARM-family and other architectures
comprising the remainder.
Compiler usage is dominated by GCC and
Clang, which together account for
approximately \UP{88\%} of compiler
occurrences; MSVC, ICC, and other compilers
constitute the rest.
Optimization settings are heavily concentrated
in standard optimization levels: O0-O3 account
for more than \UP{94\%} of their occurrences, whereas
Os, Ofast, and other appear far less frequently.
Among reversing tools, IDA Pro and Hex-Rays
(for decompilation) account for 
around \UP{$42\%$} of tool
occurrences, followed by Ghidra \UP{($19\%$)},
angr \UP{($9\%$)}, and smaller shares for
Radare2, Binary Ninja, and other tools.
The most frequently reported software packages are coreutils \UP{(19\%)} and OpenSSL \UP{(13\%)},
followed by binutils,
findutils, diffutils, BusyBox, zlib, SQLite,
curl, and others.
For binary formats, ELF accounts for 
\UP{70\%} of format occurrences, PE for
\UP{19\%}, and mixed ELF+PE or other
formats for the \UP{rest}.

\PP{Evaluation Metrics}
We organize reported evaluation metrics by
output form: classification or detection
labels, retrieval rankings, generated text,
and security-testing outcomes.
Classification and detection systems primarily
report label-agreement metrics, including
F1 (\UP{61\%}), recall (\UP{57\%}), precision (\UP{52\%}),
and accuracy (45\%).
Class-imbalanced settings additionally
report metrics such as 
macro-F1~\cite{P088_gibert2019Journal_of_Computer_Virology_and_Hacking_Techniques,
P091_malfcs2020Journal_of_Parallel_and_Distributed_Computing,
P153_snn2021saner}, while malware-drift
evaluations report time-integrated variants
such as Area Under Time~\cite{P081_TRANSCENDENT2022s&p,
P122_SCB2025ndss}.
Retrieval and similarity systems mainly
report ranking-oriented metrics: Recall@K (\UP{39\%}),
Top-K or rank (\UP{28\%}), MRR (\UP{28\%}), and
Precision@K (17\%)~\cite{P096_Genius2016ccs,
P098_vulseekerpro2018esec_fse,
P019_jtrans2022issta,P029_safe2019springer}.
Pairwise similarity formulations further report
accuracy (\UP{33\%}) and ROC-AUC (\UP{39\%}).
Generative systems for name recovery, summarization, and
pseudocode generation use a broader metric portfolio,
including precision (\UP{50\%}), recall (\UP{50\%}), F1 (\UP{44\%}),
accuracy (\UP{38\%}), BLEU (\UP{25\%}), ROUGE (\UP{19\%}),
exact match (\UP{6\%}), edit distance (\UP{6\%}),
and embedding similarity (\UP{6\%})
~\cite{P057_symlm2022ccs,P062_blens2025usenix}.
Decompilation and code-generation studies
also report functional correctness measures,
including compilability, re-executability,
pass rates, and human/tool-assisted
validation~\cite{P147_llm4decompile2024emnlp,
P149_Nova2025iclr}.
Meanwhile, security-testing systems report operational
metrics such as coverage, newly reached paths,
crashes or bugs found, solver cost, and
time-to-bug/time-to-exposure
~\cite{P131_AFLFast2016ccs,P132_neuzz2019s&p,
P133_V-Fuzz2020tcyb,P134_mtfuzz2020fse,
P135_PreFuzz2022icse,P136_AIFORE2023usenix}.
Recent LLM-based systems additionally employ
LLM-as-judge evaluation to assess generated
summaries or code explanations
~\cite{P038_ProRec2024neurips,
P068_GENNM2025ndss,P163_sk2decompile2026iclr}.

\section{Key Observations and Insights}
\label{s:insights}
\PP{Where Distinctive Binary Properties Meet AI}
\UP{The insights in this section arise from the intersection of 
AI techniques and the distinctive properties of binaries. 
AI-augmented binary reversing differs from 
generic learning settings because AI techniques operate 
on executable evidence shaped by compilation and 
conventional reverse-engineering processes: 
\PB{1}~\textit{Semantics-preserving variation}: the same 
source program can yield substantially different binaries 
across architectures, compilers, optimization levels, 
and even obfuscation schemes, while preserving semantics;
\PB{2}~\textit{Information loss}: compilation eliminates
rich source-level information, such as names, types, and
data structures;
\PB{3}~\textit{Implicit structure}: binaries do not explicitly 
encode code units, boundaries, control-flow relations, 
or data-flow relations, 
which must be recovered through additional analysis;
\PB{4}~\textit{Tool-dependent artifacts}: AI models 
learn from artifacts produced by third-party tools
(\eg disassemblers, decompilers, lifters, tracers), rendering
the resulting representations dependent 
on the behavior and assumptions of those tools;
\PB{5}~\textit{Execution-dependent behavior}: important 
semantics may emerge only under specific inputs, 
program states, or execution environments, making static artifacts
incomplete evidence for claims about runtime behavior 
or protected code; and
\PB{6}~\textit{Mixed provenance}: binaries often 
combine custom libraries, compiler- or linker-inserted code, 
and third-party components, complicating corpus construction, 
data-split policies, and provenance-sensitive evaluation.
The insights~(\IW{1}--\IW{12}) below reflect the interaction 
between these properties (\PB{1}--\PB{6}) 
and AI techniques: 
binaries determine what evidence is available, whereas 
AI pipelines determine how that 
evidence is represented, learned from, and validated.}

\subsection{Imbalances in AI-Augmented Binary Reversing}
\label{ss:rq1-landscape-maturity}
\PP{Domain Imbalance for Inference} \UP{\IW{1}}\pgap\shortlarr\pgap\UP{\PB{2}\pgap\PB{3}}
\UP{The field's uneven maturity reflects 
semantic distance from executable evidence.} 
\UP{Across the 22 domains,
bounded classification, detection, 
and retrieval domains (\UP{67.4\%}) tend to be 
mature because many of their targets are local, labels are
more readily available, and 
outputs are easier to validate. 
Code-boundary detection
~\cite{P002_xda2020ndss,P003_deepdi2022usenix,P004_tady2025usenix}
and binary similarity~\cite{P028_Gemini2017ccs,P029_safe2019springer,
P019_jtrans2022issta} exemplify this pattern: they check artifacts or paired
relations rather than broad program intent. 
Generative analyst-facing recovery (\UP{20.1\%}), 
including names, types, structures, summaries, and
pseudocode~\cite{P143_katz2018saner,P144_coda2019neurips,
P145_neutron2021Cybersecurity,P146_degpt2024ndss,
P147_llm4decompile2024emnlp}, is less mature 
because it reconstructs information that compilation 
removes or abstracts away, producing
underconstrained semantic hypotheses.
}
\UP{We cast this evidence-to-claim gap 
as a binary-specific validity issue
(\Cref{ss:rq2-validity-risks}).
}

\PP{Artifact Imbalance for Learning}\UP{\IW{2}}\pgap\shortlarr\pgap\UP{\PB{4}\pgap\PB{5}}
\UP{Artifact availability shapes apparent progress.}
\UP{
Static artifacts, such as bytes,
assembly, lifted IR, decompiled code, 
CFGs, and call graphs, dominate because
they are inexpensive to obtain and scale 
readily~\cite{P002_xda2020ndss,P028_Gemini2017ccs,
P029_safe2019springer,P019_jtrans2022issta,P147_llm4decompile2024emnlp}.
In contrast, behavior-grounded artifacts,
such as traces, memory snapshots, symbolic
constraints, and generated tests are more costly to 
collect and inherently limited by execution coverage,
yet are often necessary for analyzing 
runtime-dependent, protected, packed, or obfuscated code. 
}
Current progress therefore reflects 
what can be readily extracted and
scaled, not necessarily what best 
supports semantic validation.

\subsection{Evidence-to-Claim Validity in AI-Augmented Reversing}
\label{ss:rq2-validity-risks}

\subsubsection{Binary Corpus Validity}
\label{sss:corpus-val}
Since AI-augmented reversing 
learns from binary-derived artifacts, 
corpus construction constitutes 
a fundamental validity boundary.

\PP{Binary Corpus Monoculture}\UP{\IW{3}\pgap\shortlarr\pgap\PB{1}}
Despite the availability of 
benchmark datasets~\cite{kim2022binkit,
P019_jtrans2022issta},
most prior work relies on custom datasets, 
making direct performance comparisons difficult.
\UP{Moreover}, evaluations repeatedly 
draw from similar package families, architectures,
compilers, optimization settings, and 
build configurations, as shown 
in~\Cref{fig:plot-statistic}.
Such \emph{corpus monoculture} 
\UP{can bias evaluation toward familiar software
and build provenance, limiting generalization to
unfamiliar} architectures, toolchains, formats, 
libraries, and deployment environments.
Although increasingly large corpora 
have enabled pretraining and 
generative models~\cite{P033_palmtree2021ccs,
P036_codeart2024fse,P052_hext52023ase},
the corpus scale alone does not guarantee 
semantic generalization across
unseen samples.
This is partly because real-world 
software often incorporates
proprietary code, custom toolchains,
vendor-specific modifications, third-party
libraries, software protections, and
deployment-specific configurations that are
underrepresented in curated benchmark datasets.

\PP{Code Leakage}\UP{\IW{4}\pgap\shortlarr\pgap\PB{6}}
Train-test code leakage is a persistent threat
to validity in AI-augmented reversing
because binaries are derived from source code,
different samples may still share identical
or near-identical functions~\cite{P068_GENNM2025ndss}, libraries~\cite{P020_BinShot2022acsac,P065_dire2019ase},
templates, or compiler-generated code~\cite{P054_symgen2025ndss,P056_nfre2021issta}.
As a result, binary- or function-level dataset splits 
may inadvertently contain duplicate or 
near-duplicate code in both the training 
and evaluation sets.
Models may appear to generalize by
memorizing recurring code patterns, leading to 
overly optimistic performance estimates.
Several studies~\cite{P056_nfre2021issta,
P057_symlm2022ccs,koo2021look}
have demonstrated such leakage.
\UP{
Recent approaches mitigate such leakages
during corpus construction by removing
exact duplicates of function 
bodies~\cite{P056_nfre2021issta,
P058_asmdepictor2023ccs,P067_VARBERT2024s&p,
P140_BinT52023saner}, filtering
near-duplicates that differ by only a few tokens 
using locality-sensitive 
hashing~\cite{P147_llm4decompile2024emnlp},
or enforcing code-similarity thresholds 
between test and training 
functions~\cite{P068_GENNM2025ndss}.
}

\subsubsection{Artifact and 
Ground-Truth Fidelity}
\label{sss:artifact-val}
Analysis artifacts and ground-truth labels are 
indirect observations of program semantics, 
each introducing assumptions, inaccuracies, 
and potential sources of bias.

\PP{Proxy Label Issues}\UP{\IW{5}\pgap\shortlarr\pgap\PB{1}\pgap\PB{2}\pgap\PB{4}}
Ground truth labels in binary reversing
are often proxies rather than direct measures 
of semantic truth, each encoding 
a particular notion of correctness.
The labels 
(\eg debug symbols, compiler metadata, 
decompiler outputs, curated malware labels)
may contain inconsistencies, 
omissions, or task-specific assumptions.
\UP{Compiler optimizations complicate
downstream tasks: function
inlining~\cite{abusabha2025deep} obscures
code similarity~\cite{P022_CIDetector2024icse,P032_Asm2Vec2019sp,P015_bingo2016fse,P020_BinShot2022acsac} and 
vulnerability detection~\cite{P098_vulseekerpro2018esec_fse,P096_Genius2016ccs}
by altering function structure;
optimization-induced changes to variable
representations hinder variable-name 
recovery~\cite{P066_dirty2022usenix,P067_VARBERT2024s&p}.}
The malware analysis field
\UP{faces a different proxy-label problem due to
the scarcity of expert-verified ground truths
and the reliance on AV-derived labels}:
a study~\cite{P083_MORSE2023s&p}
reports that a widely used malware-family
dataset~\cite{joyce2023motif} 
\UP{has} approximately $40\%$ noise.
\UP{Several} works \UP{mitigate} label noise and
data sparsity~\cite{P056_nfre2021issta},
or reduce reliance on noisy labels \UP{via}
label-free representation learning
~\cite{P126_evoliot2022asia_ccs}.

\PP{Dependence on Reversing Tools}\UP{\IW{6}\pgap\shortlarr\pgap\PB{4}}
\UP{Every AI model inherits} the assumptions 
and failures of the reversing tools that
produce their inputs. 
Function boundaries, call graphs, 
decompiler variables, lifted IR, and CFGs 
are often treated as stable model inputs, 
yet they are outputs of disassemblers, 
decompilers, lifters, and tracers, which may contain errors. 
For example, prior work reports
structural inconsistencies in widely used
disassembly tools~\cite{P004_tady2025usenix},
while another study finds substantial
discrepancies in function discovery across
binary analysis tools~\cite{koo2021look}.
At higher abstraction levels, incomplete
control-flow recovery can 
propagate errors to downstream 
analyses~\cite{P013_deepvsa2019usenix},
and decompiler outputs may constitute 
noisy approximations of 
the original source~\cite{P064_jaffe2018icpc}.
Several studies therefore employ multiple
analysis tools to mitigate tool-specific
limitations~\cite{P051_CDA2025ccs}, while
others reveal substantial variation across
disassemblers, decompilers, and architectures
in the artifacts they 
produce~\cite{P007_funprobe2023esec/fse,P067_VARBERT2024s&p}.
Therefore, reported performance often
reflects the combined behavior of 
the model and its upstream toolchain rather
than the model alone.

\subsubsection{Binary Representation Validity}
\label{sss:repr-val}
\UP{
Varying forms of binary information require 
(pre)processing and encoding for
effective representations,
determining which binary-derived evidence is preserved, 
combined, or discarded before learning.
}

\PP{Representation as Semantic Commitment}\UP{\IW{7}\pgap\shortlarr\pgap\PB{1}\pgap\PB{2}}
An artifact representation
is not neutral preprocessing 
but a semantic commitment.
Artifact selection decides 
which evidence a model can observe, 
while canonicalization 
\UP{and tokenization}
decide which differences are preserved, 
abstracted, combined, or removed.  
\UP{These choices are inherently task-dependent:
normalizing registers, addresses, constants, 
offsets, strings, or identifiers 
may suppress irrelevant variation for similarity 
or classification tasks, but can also erase
evidence needed for type recovery, 
malware-behavior analysis, or vulnerability reasoning. 
Prior work illustrates these tradeoffs: byte-to-image
conversion can introduce ill-defined 
mappings~\cite{P089_SSTL2020jisa};
CFG feature selection can discard 
semantic information~\cite{P016_OrderMatters2020aaai};
selected features can achieve performance 
comparable to the full feature 
set~\cite{P124_dqfsa2019IEEE_Access}; and 
reducing reliance on CFG structure can 
improve robustness in 
binary similarity~\cite{P026_CFG2024usenix}.
The field still lacks principled guidance on what
evidence is sufficient for each reversing task and 
when incorporating additional artifacts ceases 
to provide meaningful semantic support.}

\PP{Representation Fusion}\UP{\IW{8}\pgap\shortlarr\pgap\PB{2}\pgap\PB{3}\pgap\PB{4}}
\UP{
Representation fusion raises a unique validity
question: whether multiple artifact views 
provide complementary semantic evidence or 
merely increase model complexity 
with little semantic support. 
Several systems report stronger performance 
by combining representations than
by using any one alone, including 
multi-graph embeddings~\cite{P035_guo2022icpc}, 
ensembles of pretrained instruction embeddings~\cite{P062_blens2025usenix}, 
joint code-data-address 
embeddings~\cite{P012_neudep2022esec/fse}, 
structural-and-symbol embeddings~\cite{P056_nfre2021issta}, and 
block-, string-, and import-level similarity embeddings~\cite{P095_vulhawk2023ndss}.
These suggest that complementary artifact 
views can be beneficial, but they do not 
establish a general principle that 
richer representations are inherently better. 
Representation fusion should therefore 
be evaluated through ablations against 
simpler evidence sets and justified on 
a task-specific basis according to 
how each artifact view contributes to 
the semantics being modeled.
}

\subsubsection{Model Attribution and Generalization Validity}
\label{sss:model-val}
\UP{Model validity concerns whether reported gains 
come from the claimed learning
contribution, transfer across tasks and 
binary-analysis pipelines, and remain
robust under other settings.}

\PP{Attributing Model Gains}\UP{\IW{9}\pgap\shortlarr\pgap\PB{1}\pgap\PB{2}\pgap\PB{4}}
\UP{A learning model's validity depends on whether reported gains can be attributed
to the claimed learning contribution. In AI-augmented reversing, that attribution
is difficult because performance may depend on the pretraining objective,
transferred backbone, fine-tuning recipe, representation fusion, and tuning
budget. Pretraining objectives, for example, are often presented either as
task-specific designs or as general binary-code understanding. Among designs
that claim generality, validation breadth varies: some papers test across
several downstream tasks~\cite{P033_palmtree2021ccs,P012_neudep2022esec/fse},
whereas others test essentially one, similarity
matching~\cite{P019_jtrans2022issta,P021_TREX2022tse,P027_EBM2025NeurIPS}.
Yet an independent re-evaluation finds that a vanilla BERT baseline performs
comparably to, or better than, several specialized models across tested
applications, and concludes that fine-tuning improvements are often more useful
than new pretraining tasks or architectural modifications~\cite{li2025effectiveness}}.

\PP{Model Transferability}\UP{\IW{10}\pgap\shortlarr\pgap\PB{1}\pgap\PB{2}\pgap\PB{4}}
\UP{
Model transferability 
concerns whether a pretrained backbone 
captures binary knowledge that remains useful
beyond the task, corpus, artifact form, or 
toolchain on which it was originally trained. 
Many studies reuse pretrained backbones
across relevant applications, 
including binary-code representation
models~\cite{P032_Asm2Vec2019sp,P029_safe2019springer,
P033_palmtree2021ccs}
and code LLMs~\cite{wang2021codet5,roziere2023codellama,guo2024deepseek}. 
These backbones are often frozen or adapted through
parameter-efficient fine-tuning, yet downstream
accuracy alone rarely reveals whether performance 
gains stem from transferable pretraining, 
task-specific fine-tuning data, prediction heads,
prompts, or evaluation design. 
When transfer is examined explicitly, 
reused code LLMs show relatively consistent 
gains across the studied
settings~\cite{P027_EBM2025NeurIPS,P038_ProRec2024neurips}, 
whereas smaller binary-code representation models provide 
less consistent evidence of transfer
benefits~\cite{P023_binaug2024icse,P059_xfl2023sp}. 
Claims of general
binary-code understanding hence require 
transfer assessment across downstream
domains, artifact forms, architectures, toolchains, 
and label regimes.
}

\PP{Model Robustness}\UP{\IW{11}\pgap\shortlarr\pgap\PB{1}\pgap\PB{4}\pgap\PB{5}}
The robustness of a model remains undercharacterized
both during training and after deployment.
During training, robustness depends on
ISA and platform diversity,
compiler and
optimization diversity, artifact
perturbations, 
and exposure to adversarial, obfuscated,
or otherwise challenging samples~\cite{uhm2026fool}.
Even when trained on the same corpus,
learning outcomes may vary across random
initializations, optimization trajectories,
and training configurations,
\UP{leading to} unstable performance,
overfitting, or shortcut learning 
over superficial artifact cues
rather than task-relevant program semantics.
After deployment, models 
may face \emph{distribution shift} 
from temporal drift, new \UP{analysis-}tool
versions, malware evolution, changed
compiler idioms, target architecture \UP{shifts}, and
unfamiliar execution environments.
These shifts alter both the observable
artifacts and their relationship to labels, 
degrading performance
even when the \UP{reversing} task is \UP{unchanged}.
Malware-drift 
studies~\cite{P087_transcend2017usenix,P081_TRANSCENDENT2022s&p,P126_evoliot2022asia_ccs,P162_maltree2026icml}
demonstrate that post-training degradation
is a recurring deployment risk.
\UP{Early studies}~\cite{P087_transcend2017usenix,
P081_TRANSCENDENT2022s&p} show that excluding
malware test samples associated with
low-confidence predictions \UP{can help
reduce the impact of uncertain
classifications.}
\UP{Therefore}, robustness should be evaluated
as a \emph{distributional property} across time,
toolchains, architectures, software
ecosystems, and adversarial conditions,
rather than solely through 
in-distribution test accuracy.

\subsubsection{Evaluation Validity}
\label{sss:eval-val}
Performance relies on evaluation metrics;
however, such metrics provide only indirect
evidence of semantic recovery and may fail
to capture behavioral correctness 
or analyst utility.

\PP{Metric Validity}\UP{\IW{12}\pgap\shortlarr\pgap\PB{2}\pgap\PB{5}}
Performance in AI-augmented reversing is
typically assessed through quantitative
metrics such as accuracy, precision, recall,
F1, AUC, MRR, Recall@k, BLEU, and ROUGE.
These metrics are appropriate for detection,
classification, retrieval, and generation
tasks, yet they provide only indirect
evidence of semantic recovery.
A model may achieve high retrieval accuracy
without improving analyst utility, or high
classification accuracy without capturing
causal program behavior.
\UP{For generative and source-like recovery
tasks, lexical metrics may misalign 
with semantic equivalence or human
judgment~\cite{P038_ProRec2024neurips,P161_idioms2026ndss}:
meaning-preserving summaries can receive 
zero BLEU-4~\cite{P140_BinT52023saner}, and function
names that differ only lexically may be counted as
incorrect~\cite{P054_symgen2025ndss}.}
Domain-specific evaluations partly address
this limitation.
Security-testing %
systems~\cite{P131_AFLFast2016ccs,
P132_neuzz2019s&p,
P133_V-Fuzz2020tcyb,
P134_mtfuzz2020fse,
P135_PreFuzz2022icse,
P136_AIFORE2023usenix} 
report operational
signals such as coverage, new paths,
crashes, and time-to-bug. 
Similarly, generative systems increasingly
report compilability, re-executability,
pass rates, and human- or tool-assisted
validation in addition to lexical
similarity metrics~\cite{P139_CP-BCS2023emnlp,
P140_BinT52023saner,
P141_bin2summary2024fse,
P142_misum2025fse,
P146_degpt2024ndss,
P147_llm4decompile2024emnlp,P163_sk2decompile2026iclr}.
Nevertheless, performance may still rely
on thresholds, decision boundaries,
prompts, decoding parameters, and other
evaluation choices.
What remains largely missing is a consistent
chain from model outputs to validated
semantic correctness and practical
reversing value.

\section{Open Challenges and Future Directions}
\label{sec:future}

\PP{Toward End-to-End Binary Comprehension}
\UP{Despite binary reversing being 
a family of interdependent program-semantic 
inference problems, existing AI-augmented 
binary reversing predominantly addresses 
individual reversing domains in isolation.
Hence, the next step beyond task-specific approaches is 
\emph{end-to-end binary comprehension}, 
which treats these dependencies and accumulated 
semantic evidence as a unified reasoning problem.
Achieving this goal requires models that progressively 
integrate heterogeneous binary evidence and intermediate 
semantic inferences, allowing results from one reversing 
task to constrain, refine, and validate others toward 
coherent program-level understanding.
We indicate how each insight 
(\IW{1}--\IW{12}) maps to 
the future directions discussed below.
\autoref{fig:agentic-architecture} in Appendix
illustrates a conceptual example of a future AI-augmented
binary reversing framework.
}

\PP{Representativeness-Aware Evaluation Framework}\IW{3}\ib\IW{4}\ib\IW{6}\ib\IW{9}\ib\IW{10}\ib\IW{11}
\UP{Prior AI-augmented binary reversing commonly 
demonstrates improvements on predefined corpora, 
yet aggregate corpus size and provenance provide 
limited insight into the binary properties that 
govern task difficulty. 
A corpus may differ substantially in program scale, 
control-flow complexity, indirect control transfers, 
library composition, or other structural characteristics, 
while semantics-preserving build choices further vary 
binaries across architectures, compilers, optimization 
configurations, and obfuscations. 
Consequently, performance on a predefined 
corpus provides limited evidence of generalization 
unless the represented binary population is characterized.
Future evaluation should therefore standardize 
corpus characterization along three aspects: 
\WC{1}~binary features, describing code and 
data properties of the resulting executables
(\eg program scale, function statistics, 
transformation characteristics like inlining,
obfuscation, or packing); 
\WC{2}~build provenance, describing how those binaries 
were produced (\eg architecture, source, compiler and linker information, flags, optimizations, 
libraries and components); and 
\WC{3}~analysis provenance, describing the tools and 
configurations for deriving model inputs
(\eg function discovery depends on the tool's
heuristics, and available reference artifacts). 
The latter is particularly important 
because disassemblers, 
decompilers, lifters, and tracers impose 
their own analysis policies and limitations 
on the evidence exposed to learning.
Building on such characterization, 
a common evaluation framework should report performance 
across binary properties and controlled shifts. 
This would move evaluation beyond aggregate performance 
on a fixed corpus toward establishing 
where a technique is valid, what binary population 
it represents, and under which build 
and analysis conditions its conclusions generalize.}

\PP{Binary-Aligned Semantic Ground Truth}\IW{1}\pgap\IW{5}\pgap\IW{12}
\UP{
Many binary reversing tasks enrich 
decompiled code with source-level semantics, including 
symbols, types, data structures, and higher-level abstractions. 
However, their evaluation often relies on 
the original source as ground truth. 
This correspondence becomes increasingly 
ill-defined under compiler transformations: inlining 
removes standalone function boundaries, dead-code 
elimination removes source constructs entirely, 
vectorization changes computational structure, and 
compiler- or linker-generated code may have 
no direct source counterpart. 
Consequently, requiring recovered binary semantics 
to reproduce the original source can conflate 
reversing errors with irreversible compilation effects.
A promising direction is to construct 
\emph{binary-aligned semantic ground truth}
during compilation. 
For instance, compiler-maintained information, including 
post-optimization intermediate representations, 
debugging metadata, and optimization provenance,
could be used to 
derive a C-like semantic representation after 
transformations have taken effect. 
Such representations would provide supervision 
aligned with the executable observed by 
the reversing system while retaining source-derived 
semantic information available during compilation.
This would shift evaluation from asking 
whether a model reconstructs the original source 
toward whether it recovers the highest-level semantics 
justified by the resulting binary, providing 
a more principled target for decompilation 
and downstream semantic inference.
}

\PP{Multi-step Semantic Reasoning}\IW{1}\pgap\IW{12}
\UP{
Multi-step reasoning remains relatively 
underexplored in AI-augmented binary reversing.
It refers to 
Chain-of-Thought (CoT)-style reasoning, 
in which intermediate semantic inferences 
are generated to support 
subsequent conclusions.
Recent studies~\cite{zhang2026coder, liu2026binary}
have begun to investigate rationale-guided generation
and iterative feedback-driven refinement.
While current AI approaches 
commonly regard inference tasks as isolated predictions,
future AI-augmented systems should 
reason through intermediate semantic hypotheses, 
allowing information recovered at one step 
to constrain, refine, and validate 
subsequent inferences. 
Such reasoning may involve problem decomposition, 
iterative refinement, hypothesis testing, and 
feedback from analysis tools, 
rather than relying on a single prediction step.}

\PP{Multimodal Semantic Reasoning}\IW{2}\pgap\IW{7}\pgap\IW{8}
\UP{Multimodal reasoning addresses the incompleteness of 
any single source of binary evidence, while 
multi-step reasoning addresses dependencies
among semantic inference tasks.
Rather than treating static, dynamic, and triage information 
as independent feature sources to be fused into 
a single representation, future AI-augmented reversing
systems should organize them as complementary sources
of semantic evidence (\eg  unknown vulnerability 
discovery~\cite{P160_firmagent2026ndss}).}
\UP{
Initial triage can establish hypotheses and 
prioritize analysis, while static, dynamic, symbolic,
memory, and execution-based analyses can be 
selectively acquired to resolve uncertainty, 
test hypotheses, and refine intermediate 
semantic inferences.
The challenge is therefore not merely to fuse 
heterogeneous artifacts, but to determine which evidence 
is needed for a given semantic claim and to reconcile 
potentially complementary or conflicting observations 
while preserving their provenance and uncertainty.}

\PP{Goal-Driven Binary Reversing with Agentic AI}\IW{2}\pgap\IW{6}\pgap\IW{12}
\UP{Multi-step and multimodal semantic reasoning provide the
foundation for binary reversing with agentic AI: the former
structures intermediate semantic inferences, while the latter
broadens the evidence available to support them.
Agentic AI can extend these capabilities toward
goal-driven analysis by deciding which reasoning step
to pursue, what evidence to acquire, and which analysis
tool to invoke next.
As an example, given a task that determines 
whether a packed binary performs credential theft, 
an agent could form an initial analysis plan 
from triage information, retrieve
relevant knowledge about the 
suspected packer or APIs, invoke 
dynamic analysis to expose unpacked code, and
then use a debugger to set breakpoints at
suspicious routines.
The agent could then inspect execution traces,
register values, and memory states for
credential-related behavior, using newly acquired
evidence to refine or reject its hypotheses
and determine subsequent analysis steps.
Hence, such systems would move beyond fixed analysis
pipelines toward adaptive and closed-loop 
reversing workflows in which
planning, reasoning, retrieval, 
evidence acquisition, and tool orchestration
are jointly guided by the semantic objective.}

\PP{AI-Powered Autonomous Binary 
Reversing and Human Roles}\IW{6}\pgap\IW{12}
\UP{The evolution from AI-augmented to
AI-powered reversing should be viewed as
an autonomy spectrum rather than
a replacement of human analysts.
As agents assume greater responsibility for executing
reversing workflows, the central question becomes which
analytical decisions can be delegated and which should
remain subject to human judgment.
For instance, an agent may autonomously recover a call chain,
infer that user-controlled data reaches a memory operation,
and assemble evidence suggesting a vulnerability; however,
a human analyst may still need to judge whether the behavior
is practically exploitable, whether environmental assumptions
are realistic, and whether the evidence is sufficient to
support a security-critical conclusion.
A recent empirical study~\cite{basque2026decompiling} 
on human-LLM teaming shows that LLM assistance 
can improve the performance of novices, but
provide limited benefits to experts, or even mislead them 
through hallucinations.
}
\UP{
Autonomous binary reversing systems 
should therefore expose the provenance,
uncertainty, and supporting evidence behind 
their conclusions,
while humans retain responsibility 
for defining objectives,
validation criteria, and consequential decisions.
Establishing appropriate autonomy
boundaries is an open challenge: 
which reversing activities can be delegated
reliably, which conclusions require explicit supporting
evidence, and where human review or intervention remains
necessary.}

\section{Concluding Remarks}
\label{sec:conclusion}
\UP{We present the first systematization of 
AI-augmented binary reversing, covering 
246 papers across 22 domains. 
We introduce a unified taxonomy that bridges 
conventional and AI-augmented binary reversing 
using analysis artifacts as the interface 
between reversing workflows and AI-driven inference.}
\UP{
Our in-depth analysis demonstrates that 
progress in this field relies not only on 
stronger models, but also on 
how binary evidence is produced, represented, 
combined, and validated.
Rather than treating reversing tasks 
and artifact sources independently, 
future AI-powered binary reversing systems 
should progressively accumulate, reconcile, 
and validate semantic evidence across tasks, 
modalities, tools, and execution contexts 
to construct coherent program-level understanding.
We envision AI-augmented binary reversing
evolving from task-specific prediction 
toward \emph{end-to-end binary comprehension}, 
supported by representative evaluation, 
binary-aligned semantic ground truth,
multi-step and multimodal semantic reasoning, 
and goal-driven agentic analysis.
}

\section*{Disclosure of 
Generative AI Usage}
\noindent
Generative AI tools
were employed during 
the literature review process 
to help retrieve information 
from the collected papers according
to the proposed taxonomy
and identify 
potentially relevant content 
for further examination. 
However, all AI-generated outputs 
were treated 
as auxiliary aids and  
manually verified against the original papers by the authors. 
Every technical claim, 
taxonomy design, 
categorization of prior work,
trend analysis, 
synthesis of findings, and 
scientific conclusions were
developed, reviewed, and
approved by the authors,
who take full responsibility 
for the accuracy and
integrity of this manuscript.

\balance
\footnotesize
\bibliographystyle{IEEEtran}
\bibliography{references}

\appendix

\normalsize

\begin{figure*}[!t]
    \centering
    \includegraphics[width=\linewidth]{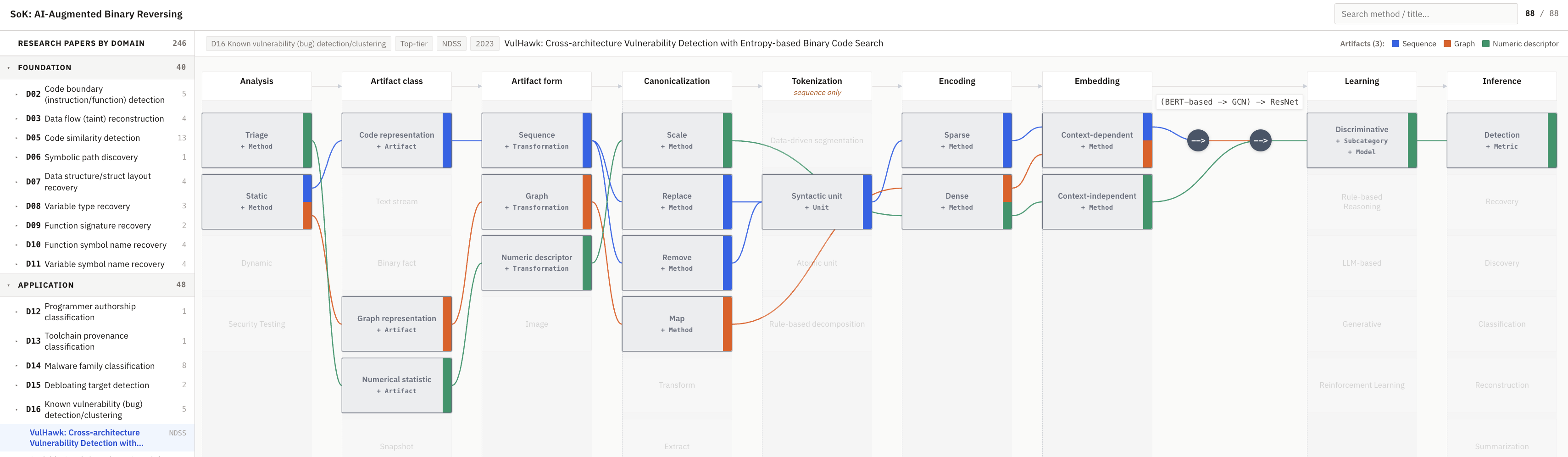}
    \caption{Snapshot of our interactive visualization platform for AI-augmented binary reversing pipelines.
    The left panel groups papers by reversing domain, listing each domain's share of the corpus and links to its papers, while the main panel summarizes a selected study across the pipeline stages: artifact selection (\ie analysis mode, artifact class and form), canonicalization, tokenization, encoding, embedding, learning, and inference.
    Each color denotes a model-consumable artifact form, and its connections indicate the flow across subsequent stages.
    Clicking a box displays additional information if available.}
    \label{fig:web-example}

    \vspace{2.0em}

    \begin{minipage}[!t]{0.48\textwidth}
        \centering
        \includegraphics[width=0.95\linewidth]{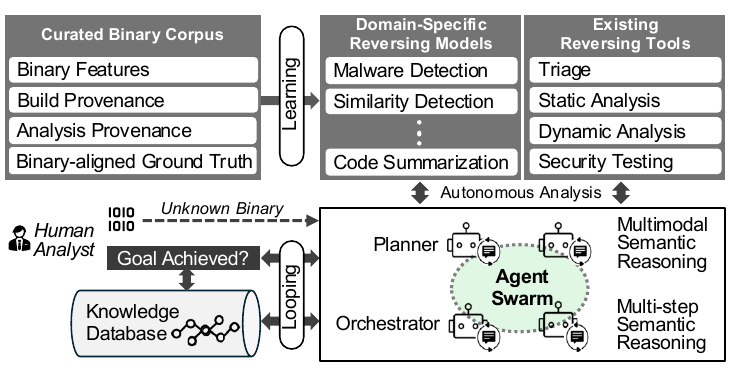}
        \caption{
         \UP{
        A conceptual example of an AI-augmented
        binary reversing framework for
        end-to-end binary comprehension.
        Domain-specific AI models (\eg malware detection)
        are prepared using curated binary corpora
        enriched with binary features, build and analysis
        provenance, and binary-aligned ground truth.
        Given an unknown binary, a human analyst
        specifies a goal, and an agent swarm
        autonomously plans, orchestrates, and
        performs the analysis using existing reversing
        tools and domain-specific AI models as needed.
        The autonomous analysis involves multi-step and
        multimodal semantic reasoning.
        The analysis proceeds iteratively, with the analyst
        evaluating whether the goal has been achieved and
        the framework accumulating knowledge in
        a knowledge database until the goal is satisfied.
        }
        }
        \label{fig:agentic-architecture}
    \end{minipage}%
    \hfill
    \begin{minipage}[t]{\columnwidth}
        \hfill
    \end{minipage}
\end{figure*}

\begin{table*}[htb]\centering\scriptsize
\caption{
Proposed AI-augmented binary reversing 
pipeline via analysis artifacts, as
observed across \UP{88} top-venue papers.
For each study, this table summarizes
how its analysis artifacts traverse 
the pipeline, including artifact class (form) 
canonicalization, tokenization, encoding, 
embedding, and learning paradigm
(\Cref{subsec:artifact-forms}--\Cref{subsec:learning-paradigms}). 
Notably, different artifact forms 
(\Cref{tbl:analysis-artifact-taxonomy}) 
follow distinct canonicalization, 
tokenization, encoding, 
and embedding pathways.
Within a cell, semicolons (;)
separate multiple model inputs and
em dashes (---) indicate 
information unavailable.
The symbols \protect\iconA,
\protect\iconS, \protect\iconR,
and \protect\iconD~denote atomic,
syntactic, rule-based, and data-driven
tokenization, respectively~(\Cref{tab:tok-taxonomy});
\protect\embctx, \protect\embna, and
\protect\ding{55} denote context-dependent,
context-independent,
and absent embeddings, respectively~(\Cref{tbl:encoding-embedding-taxonomy}).
At the embedding stage,
$\rightarrow$ denotes sequential
embedding, where a later embedding is
initialized from a preceding one,
whereas $\parallel$ denotes
embedding concatenation.
}
  \label{tab:ai-pipeline}
  \resizebox{\textwidth}{!}{
  \fontsize{9}{10}\selectfont
  \begin{tabular}{@{}c|l|l|l|l|l@{}}
\hline
\textbf{DID} 
& \multicolumn{1}{c|}{\textbf{Study}}
& \multicolumn{1}{c|}{\textbf{Artifact form (Artifact class)}}
& \multicolumn{1}{c|}{\textbf{Canonicalization (Tokenization)}}
& \multicolumn{1}{c|}{\textbf{Encoding (Embedding)}}
& \multicolumn{1}{c}{\textbf{Learning}}\\
\hline
D02 & Shin et al.~\cite{P005_shin2015usenix}
&\shortstack{\iconseq~(Code)}
&\shortstack{---~(\iconA)}
&\shortstack{Sparse~(\embctx)}
&\shortstack{Discriminative}\\
 & XDA~\cite{P002_xda2020ndss}
&\shortstack{\iconseq~(Code)}
&\shortstack{---~(\iconA)}
&\shortstack{Sparse~(\embctx)}
&\shortstack{Discriminative}\\
 & DeepDi~\cite{P003_deepdi2022usenix}
&\shortstack{\iconnum~(Code);~\icongraph~(Graph)}
&\shortstack{Remove; map}
&\shortstack{Dense~(\embctx)~$\rightarrow$~Dense~(\embctx)}
&\shortstack{Discriminative}\\
 & FunProbe~\cite{P007_funprobe2023esec/fse}
&\shortstack{\icongraph~(BinFact,~Code,~Graph)}
&\shortstack{Transform}
&\shortstack{Dense~(\protect\ding{55})}
&\shortstack{Rule-based}\\
 & Tady~\cite{P004_tady2025usenix}
&\shortstack{\iconseq~(Code);~\iconseq(Graph)}
&\shortstack{---~(\iconA); ---}
&\shortstack{Sparse~(\embctx)~$\rightarrow$~Dense~(\embctx)}
&\shortstack{Discriminative}\\
\hline
D03 & TAINTINDUCE~\cite{P011_TAINTINDUCE2019ndss}
&\shortstack{\iconnum~(BinFact, Snapshot)}
&\shortstack{---}
&\shortstack{Dense~(\protect\ding{55})}
&\shortstack{Rule-based}\\
 & DEEPVSA~\cite{P013_deepvsa2019usenix}
&\shortstack{\iconseq~(TextStream)}
&\shortstack{Scale~(\iconA)}
&\shortstack{Sparse~(\embctx)}
&\shortstack{Discriminative}\\
 & Neutaint~\cite{P008_neutaint2020sp}
&\shortstack{\iconnum~(BinFact,~TestSet)}
&\shortstack{Scale}
&\shortstack{Dense~(\protect\ding{55})}
&\shortstack{Discriminative}\\
 & NeuDep~\cite{P012_neudep2022esec/fse}
&\shortstack{\iconseq~(TextStream);~\iconseq~(BinFact);~\iconseq~(TextStream)}
&\shortstack{Remove~(\iconS);~scale~(\iconA);~scale~(\iconA)}
&\shortstack{Sparse~(\embctx)~$\parallel$~Dense~(\embna)~$\parallel$~Dense~(\embna)}
&\shortstack{Discriminative}\\
\hline
D05 & Gemini~\cite{P028_Gemini2017ccs}
&\shortstack{\icongraph~(Code,~Graph,~TextStream,~BinFact)}
&\shortstack{Map}
&\shortstack{Dense~(\embctx)}
&\shortstack{Discriminative}\\
 & Li et al.~\cite{P030_li2019pmlr}
&\shortstack{\icongraph~(Code,~Graph)}
&\shortstack{Map}
&\shortstack{Dense~(\embctx)}
&\shortstack{Discriminative}\\
 & INNEREYE~\cite{P031_INNEREYE2019ndss}
&\shortstack{\iconseq~(Code)}
&\shortstack{Replace~(\iconS)}
&\shortstack{Sparse~(\embctx)}
&\shortstack{Discriminative}\\
 & Asm2Vec~\cite{P032_Asm2Vec2019sp}
&\shortstack{\iconseq~(Code,~Graph)}
&\shortstack{Replace~(\iconS)}
&\shortstack{Sparse~(\embna)}
&\shortstack{Discriminative}\\
 & CI-Detector~\cite{P022_CIDetector2024icse}
&\shortstack{\icongraph~(Code,~Graph)}
&\shortstack{---}
&\shortstack{Dense~(\embctx)}
&\shortstack{Discriminative}\\
 & HermesSim~\cite{P024_HermesSim2024usenix}
&\shortstack{\icongraph~(Code,~Graph)}
&\shortstack{Map,~transform}
&\shortstack{Sparse~(\embctx)}
&\shortstack{Discriminative}\\
 & EBM~\cite{P027_EBM2025NeurIPS}
&\shortstack{\iconseq~(Code)}
&\shortstack{Replace~(\iconD)}
&\shortstack{Sparse~(\embctx)}
&\shortstack{LLM-based}\\
 & BINGO~\cite{P015_bingo2016fse}
&\shortstack{\iconnum~(BinFact,~Code,~Graph,~LogicExp)}
&\shortstack{---}
&\shortstack{Sparse~(\protect\ding{55})}
&\shortstack{Rule-based}\\
 & OrderMatters~\cite{P016_OrderMatters2020aaai}
&\shortstack{\iconseq~(Code);~\icongraph~(Graph)}
&\shortstack{---~(\iconD);~---}
&\shortstack{Sparse~(\embctx)~$\rightarrow$~Dense~(\embctx)}
&\shortstack{Discriminative}\\
 & BINAUG~\cite{P023_binaug2024icse}
&\shortstack{\iconseq~(Code);~\icongraph~(Graph)}
&\shortstack{Replace~(\iconS);~---}
&\shortstack{Sparse~(\embctx)~$\rightarrow$~Dense~(\embctx)}
&\shortstack{Discriminative}\\
 & BinaryAI~\cite{P025_binaryai2024icse}
&\shortstack{\iconseq~(Code)}
&\shortstack{---~(\iconD)}
&\shortstack{Sparse~(\embctx)}
&\shortstack{LLM-based}\\
 & CodeCMR~\cite{P017_codecmr2020nips}
&\makecell[l]{\iconseq~(Code);~\icongraph~(Graph);~\iconseq~(TextStream);~\iconseq~(BinFact)}
&\makecell[l]{---;~---;~---~(\iconA);~---}
&\makecell[l]{(Sparse~(\embctx)~$\rightarrow$~Dense~(\embctx))~$\parallel$ \\
Sparse~(\embctx)~$\parallel$~Sparse~(\embctx)}
&\shortstack{Discriminative}\\
 & \UP{BDLF-Qwen3~\cite{P156_bdlfqwen32026AAAI}}
&\makecell[l]{\iconseq~(Code, Code);~\iconseq~(Code, Code)}
&\makecell[l]{Replace~(\iconD); Replace~(\iconD)}
&\makecell[l]{Sparse~(\embctx)~$\rightarrow$~Sparse~(\embctx)}
&\shortstack{Discriminative}\\
\hline
D06 & Learch~\cite{P040_Learch2021CCS}
&\shortstack{\iconnum~(BinFact)}
&\shortstack{---}
&\shortstack{Dense~(\protect\ding{55})}
&\shortstack{Discriminative}\\
\hline
D07 & OSPREY~\cite{P041_osprey2021s&p}
&\shortstack{\icongraph~(BinFact,~Code,~Graph)}
&\shortstack{Transform}
&\shortstack{Dense~(\protect\ding{55})}
&\shortstack{Rule-based}\\
 & ReSym~\cite{P042_resym2024ccs}
&\makecell[l]{\iconseq~(BinFact,~Code);~\iconseq~(BinFact,~Code);
\icongraph~(BinFact,~Graph)}
&\makecell[l]{Replace~(\iconD);~Replace~(\iconD);~Transform}
&\makecell[l]{Sparse~(\embctx)~$\parallel$~Sparse~(\embctx);~Dense~(\protect\ding{55})}
&\shortstack{LLM-based; Rule-based}\\
 & TypeForge~\cite{P043_typeforge2025sp}
&\shortstack{\icongraph~(BinFact,~Code,~Graph);~\iconseq~(Code)}
&\shortstack{Map,~transform;~---~(\iconD)}
&\shortstack{Dense~(\protect\ding{55});~Sparse~(\embctx)}
&\shortstack{Rule-based;~LLM-based}\\
& \UP{CiRCLE~\cite{P159_circe2026SP}}
&\shortstack{\icongraph~(BinFact,~Code,~Graph)}
&\shortstack{Map}
&\shortstack{Dense~(\protect\ding{55})}
&\shortstack{Rule-based;~LLM-based}\\
\hline
D08 & TYGR~\cite{P048_tygr2024usenix}
&\shortstack{\icongraph~(Code,~Graph)}
&\shortstack{Transform}
&\shortstack{Sparse~(\embctx)}
&\shortstack{Discriminative}\\
 & TRex~\cite{P049_TRex2025usenix}
&\shortstack{\icongraph~(BinFact,~Code,~Graph)}
&\shortstack{Transform}
&\shortstack{Dense~(\protect\ding{55})}
&\shortstack{Rule-based}\\
 & StateFormer~\cite{P044_stateformer2021fse}
&\shortstack{\iconseq~(BinFact,~Code,~TextStream);~\iconseq~(Code)}
&\shortstack{Replace~(\iconS);~Replace~(\iconS)}
&\shortstack{Sparse~(\embctx)~$\rightarrow$~Sparse~(\embctx)}
&\shortstack{Discriminative}\\
\hline
D09 & EKLAVYA~\cite{P050_EKLAVYA2017usenix}
&\shortstack{\iconseq~(Code,~Graph);~\iconseq~(Code)}
&\shortstack{---~(\iconS);~---~(\iconS)}
&\shortstack{Sparse~(\embna)~$\rightarrow$~Sparse~(\embctx)}
&\shortstack{Discriminative}\\
 & CDA~\cite{P051_CDA2025ccs}
&\shortstack{\icongraph~(Code,~Graph)}
&\shortstack{Transform}
&\shortstack{Dense~(\protect\ding{55})}
&\shortstack{Rule-based}\\
\hline
D10 & XFL~\cite{P059_xfl2023sp}
&\shortstack{\iconnum~(Graph);~\iconnum~(Graph);~\iconnum~(BinFact,~Code)}
&\shortstack{---;~---;~Scale}
&\shortstack{(Dense~(\embna)~$\parallel$~Dense~(\embna))~$\rightarrow$~Dense~(\embna)}
&\shortstack{Discriminative}\\
 & SYMGEN~\cite{P054_symgen2025ndss}
&\shortstack{\iconseq~(Code)}
&\shortstack{Replace~(\iconD)}
&\shortstack{Sparse~(\embctx)}
&\shortstack{LLM-based}\\
 & SymLM~\cite{P057_symlm2022ccs}
&\shortstack{\iconseq~(TextStream);~\iconnum~(Graph)}
&\shortstack{Replace~(\iconS);~---}
&\shortstack{Sparse~(\embctx)~$\parallel$~Dense~(\embna)}
&\shortstack{Discriminative}\\
 & BLens~\cite{P062_blens2025usenix}
&\shortstack{\iconseq~(Code);~\iconseq~(Code);~\iconnum~(BinFact,~Code,~Graph)}
&\shortstack{Replace~(\iconD);~Replace~(\iconD);~Scale}
&\shortstack{Sparse~(\embctx)~$\parallel$~Sparse~(\embctx)~$\parallel$~Dense~(\embna)}
&\shortstack{Generative}\\
\hline
D11 & VARBERT~\cite{P067_VARBERT2024s&p}
&\shortstack{\iconseq~(Code)}
&\shortstack{Replace~(\iconD)}
&\shortstack{Sparse~(\embctx)}
&\shortstack{Discriminative}\\
 & Debin~\cite{P063_debin2018ccs}
&\shortstack{\iconnum~(BinFact,~Code,~Graph,~TextStream);~\icongraph~(Graph)}
&\shortstack{---;~Map}
&\shortstack{Sparse~(\protect\ding{55});~Sparse~(\protect\ding{55})}
&\shortstack{Discriminative; Rule-based}\\
 & DIRTY~\cite{P066_dirty2022usenix}
&\shortstack{\iconseq~(Code);~\iconseq~(BinFact)}
&\shortstack{Replace~(\iconD);~Replace~(\iconR)}
&\shortstack{Sparse~(\embctx)~$\parallel$~Sparse~(\embctx)}
&\shortstack{Generative}\\
 & GENNM~\cite{P068_GENNM2025ndss}
&\shortstack{\iconseq~(Code,~Graph)}
&\shortstack{Replace~(\iconD)}
&\shortstack{Sparse~(\embctx)}
&\shortstack{LLM-based}\\
\hline
D12 & Caliskan et al.~\cite{P070_caliskan2015ndss}
&\makecell[l]{\iconnum~(BinFact,~Code,~TextStream);~\iconnum~(Graph);\\
\iconnum~(Code);~\iconnum~(Graph)}
&\makecell[l]{---;~---;~---;~---}
&\makecell[l]{Dense~(\protect\ding{55})~$\parallel$~Dense~(\protect\ding{55})~$\parallel$\\
Dense~(\protect\ding{55})~$\parallel$~Dense~(\protect\ding{55})}
&\shortstack{Discriminative}\\
\hline
D13 & Du et al.~\cite{P075_du_extend2023ccs}
&\shortstack{\iconnum~(Code)}
&\shortstack{---}
&\shortstack{Sparse~(\embna)}
&\shortstack{Discriminative}\\
\hline
D14 & Transcend~\cite{P087_transcend2017usenix}
&\shortstack{\iconnum~(Code)}
&\shortstack{---}
&\shortstack{Dense~(\protect\ding{55})}
&\shortstack{Discriminative}\\
 & DeepReflect~\cite{P151_deepreflect2021usenix}
&\shortstack{\iconnum~(Code,~Graph)}
&\shortstack{Scale}
&\shortstack{Dense~(\embna)}
&\shortstack{Discriminative}\\
 & Gibert et al.~\cite{P082_gibert2018aaai}
&\shortstack{\iconseq~(Numeric)}
&\shortstack{---~(\iconR)}
&\shortstack{Dense~(\protect\ding{55})}
&\shortstack{Discriminative}\\
 & MORSE~\cite{P083_MORSE2023s&p}
&\shortstack{\iconnum~(BinFact,~Numeric,~TextStream)}
&\shortstack{Scale}
&\shortstack{Dense~(\protect\ding{55})}
&\shortstack{Discriminative}\\
 & EMBERSim~\cite{P150_embersim2023NeurIPS}
&\shortstack{\iconnum~(BinFact,~Numeric,~TextStream)}
&\shortstack{---}
&\shortstack{Dense~(\protect\ding{55})}
&\shortstack{Discriminative}\\
 & MalCL~\cite{P084_malcl2025aaai}
&\shortstack{\iconnum~(Numeric,~BinFact)}
&\shortstack{Scale}
&\shortstack{Dense~(\protect\ding{55})}
&\shortstack{Discriminative}\\
 & \UP{LMLMs}~\cite{P157_lmlms2026ndss}
&\shortstack{\iconseq~(Code); \iconseq~(Code); \iconseq~(Code)}
&\shortstack{---~(\iconD); ---~(\iconD); ---~(\iconD)}
&\shortstack{Sparse~(\embctx); Sparse~(\embctx); Sparse~(\embctx)}
&\shortstack{Discriminative}\\
 & \UP{MalTree~\cite{P162_maltree2026icml}}
&\shortstack{\iconnum~(Numeric,~BinFact,~TextStream);
\iconseq~(TextStream);
\iconimg~(Snapshot)
}
&\shortstack{---; Remove~(\iconD); Scale}
&\shortstack{Dense~(\embna); Sparse~(\embna); Dense~(\embna)}
&\shortstack{Discriminative}\\
\hline
D15 & CCFG~\cite{P093_CCFG2019ccs}
&\shortstack{\icongraph~(Code,~Graph,~TextStream)}
&\shortstack{Transform}
&\shortstack{Sparse~(\protect\ding{55})}
&\shortstack{Discriminative}\\
 & Picup~\cite{P094_Picup2023esec/fse}
&\shortstack{\iconimg~(TestSet);~\iconseq~(TestSet)}
&\shortstack{Scale;~---~(\iconR)}
&\shortstack{Dense~(\embctx)~$\parallel$~Dense~(\embna)}
&\shortstack{Discriminative}\\
\hline
D16 & Genius~\cite{P096_Genius2016ccs}
&\shortstack{\iconnum~(Code,~Graph)}
&\shortstack{---}
&\shortstack{Dense~(\protect\ding{55})}
&\shortstack{Discriminative}\\
 & discovRE~\cite{P097_discovre2016ndss}
&\shortstack{\iconnum~(BinFact,~Code,~Graph)}
&\shortstack{Scale}
&\shortstack{Dense~(\protect\ding{55})}
&\shortstack{Discriminative}\\
 & Vulseeker-pro~\cite{P098_vulseekerpro2018esec_fse}
&\shortstack{\icongraph~(Code,~Graph)}
&\shortstack{Map}
&\shortstack{Dense~(\embctx)}
&\shortstack{Discriminative}\\
 & MDSAE~\cite{P103_MDSAE2019iclr}
&\shortstack{\iconseq~(Code)}
&\shortstack{Remove~(\iconS)}
&\shortstack{Sparse~(\embctx)}
&\shortstack{Discriminative}\\
 & VulHawk~\cite{P095_vulhawk2023ndss}
&\shortstack{\iconseq~(Code);~\icongraph~(Graph);~\iconnum~(Numeric)}
&\shortstack{Replace,~remove~(\iconS);~Map;~Scale}
&\shortstack{(Sparse~(\embctx)~$\rightarrow$~Dense~(\embctx))~$\rightarrow$~Dense~(\embna)}
&\shortstack{Discriminative}\\
\hline
D17 & Yu Wang et al.~\cite{P107_Wang2020aaai}
&\shortstack{\iconseq~(TextStream);~\iconnum~(TextStream)}
&\shortstack{---~(\iconS);~---}
&\shortstack{Sparse~(\embctx);~Sparse~(\protect\ding{55})}
&\shortstack{Discriminative; Reinforcement}\\
 & TRANSCENDENT~\cite{P081_TRANSCENDENT2022s&p}
&\shortstack{\iconnum~(BinFact,~Numeric,~TextStream)}
&\shortstack{Scale}
&\shortstack{Dense~(\protect\ding{55})}
&\shortstack{Discriminative}\\
  & Zhang et al.~\cite{P108_zhang2020aaai}
&\shortstack{\iconseq~(TextStream)}
&\shortstack{Remove~(\iconS)}
&\shortstack{Dense~(\embctx)}
&\shortstack{Discriminative}\\
 & PROVDETECTOR~\cite{P109_PROVDETECTOR2020ndss}
&\shortstack{\iconseq~(Graph)}
&\shortstack{Remove,~replace~(\iconR)}
&\shortstack{Sparse~(\embna)}
&\shortstack{Discriminative}\\
 & MalConv2~\cite{P110_MalConv22021aaai}
&\shortstack{\iconseq~(Code)}
&\shortstack{---~(\iconA)}
&\shortstack{Sparse~(\embna)}
&\shortstack{Discriminative}\\
 & Greedy~\cite{P113_Greedy2023usenix}
&\shortstack{\iconseq~(Code)}
&\shortstack{---~(\iconA)}
&\shortstack{Sparse~(\embna)}
&\shortstack{Discriminative}\\
 & GreedyBlock~\cite{P117_GreedyBlock2024ccs}
&\shortstack{\iconseq~(Code)}
&\shortstack{---~(\iconA)}
&\shortstack{Sparse~(\embna)}
&\shortstack{Discriminative}\\
 & Shuai Li et al.~\cite{P119_shuai2024ndss}
&\shortstack{\iconseq~(Code);~\icongraph~(Graph)}
&\shortstack{Replace~(\iconD);~Map}
&\shortstack{Sparse~(\embctx)~$\rightarrow$~Dense~(\embctx)}
&\shortstack{Discriminative}\\
 & SRDC~\cite{P120_srdc2025aaai}
&\shortstack{\iconseq~(TextStream)}
&\shortstack{Replace, remove~(\iconD)}
&\shortstack{Sparse~(\embctx)}
&\shortstack{Discriminative}\\
 & ERW-Radar~\cite{P121_ERW-Radar2025ndss}
&\shortstack{\iconseq~(TextStream);~\iconnum~(Numeric)}
&\shortstack{Remove~(\iconS);~Scale}
&\shortstack{Sparse~(\embctx);~Dense~(\protect\ding{55})}
&\shortstack{Discriminative}\\
 & SCB~\cite{P122_SCB2025ndss}
&\shortstack{\iconnum~(BinFact,~Numeric,~TextStream)}
&\shortstack{Scale}
&\shortstack{Dense~(\protect\ding{55})}
&\shortstack{Discriminative}\\
 & Adv-MalBayes~\cite{P114_Adv-MalBayes2023aaai}
&\shortstack{\iconnum~(BinFact,~Numeric,~TextStream)}
&\shortstack{---}
&\shortstack{Dense~(\protect\ding{55})}
&\shortstack{Discriminative}\\
\hline
D18 & DEEPBINDIFF~\cite{P128_deepbindiff2020ndss}
&\shortstack{\iconseq~(Code);~\icongraph~(Graph)}
&\shortstack{Replace~(\iconS);~Transform}
&\shortstack{Dense~(\embna)~$\rightarrow$~Dense~(\embctx)}
&\shortstack{Discriminative}\\
&
\UP{BINALIGNER~\cite{P158_binaligner2026ndss}}
&\shortstack{\icongraph~(Code, Graph)}
&\shortstack{Map}
&\shortstack{Dense~(\embctx)}
&\shortstack{Discriminative}\\
\hline
D19 & AFLFast~\cite{P131_AFLFast2016ccs}
&\shortstack{\iconnum~(Numeric)}
&\shortstack{---}
&\shortstack{Dense~(\protect\ding{55})}
&\shortstack{Rule-based}\\
 & NEUZZ~\cite{P132_neuzz2019s&p}
&\shortstack{\iconseq~(Numeric,~TestSet)}
&\shortstack{---~(\iconA)}
&\shortstack{Dense~(\protect\ding{55})}
&\shortstack{Discriminative}\\
 & MTFuzz~\cite{P134_mtfuzz2020fse}
&\shortstack{\iconseq~(TestSet)}
&\shortstack{Scale~(\iconA)}
&\shortstack{Dense~(\embna)}
&\shortstack{Discriminative}\\
 & FICS~\cite{P130_FICS2021usenix}
&\shortstack{\icongraph~(Code,~Graph);~\icongraph~(Code,~Graph)}
&\shortstack{Extract,~transform;~Extract,~transform}
&\shortstack{Dense~(\embna)~$\rightarrow$~Dense~(\embctx)}
&\shortstack{Discriminative}\\
 & PreFuzz~\cite{P135_PreFuzz2022icse}
&\shortstack{\iconseq~(TestSet)}
&\shortstack{Scale~(\iconA)}
&\shortstack{Dense~(\embna)}
&\shortstack{Discriminative}\\
 & UVSCAN~\cite{P101_uvscan2023usenix}
&\shortstack{\icongraph~(Code,~Graph)}
&\shortstack{Transform}
&\shortstack{Dense~(\protect\ding{55})}
&\shortstack{Rule-based}\\
 & AIFORE~\cite{P136_AIFORE2023usenix}
&\shortstack{\iconseq~(TextStream);~\iconnum~(Code,~Graph,~TextStream)}
&\shortstack{---~(\iconS);~---}
&\shortstack{Sparse~(\protect\ding{55});~Dense~(\embctx)}
&\shortstack{Rule-based;~Discriminative}\\
 & HermeScan~\cite{P104_HermeScan2024ndss}
&\shortstack{\iconseq~(TextStream);~\icongraph~(BinFact,~Code,~Graph)}
&\shortstack{---~(\iconD);~Transform}
&\shortstack{Sparse~(\embctx);~---~(\protect\ding{55})}
&\shortstack{Discriminative}\\
 & \UP{FirmAgent~\cite{P160_firmagent2026ndss}}
&\shortstack{\iconseq~(Code,~Graph,~Snapshot);~\iconseq~(Code,~TestSet)}
&\shortstack{---~(\iconD);~---~(\iconD)}
&\shortstack{Sparse~(\embctx)~$\parallel$~Sparse~(\embctx)}
&\shortstack{LLM-based; LLM-based}\\
\hline
D21 & Bin2Summary~\cite{P141_bin2summary2024fse}
&\shortstack{\iconseq~(Code,~Graph)}
&\shortstack{Replace~(\iconS)}
&\shortstack{Sparse~(\embctx)}
&\shortstack{Generative}\\
 & MiSum~\cite{P142_misum2025fse}
&\shortstack{\icongraph~(Code,~Graph)}
&\shortstack{Map,~transform}
&\shortstack{Dense~(\embctx)}
&\shortstack{Generative}\\
 & ProRec~\cite{P038_ProRec2024neurips}
&\shortstack{\iconseq~(Code);~\icongraph~(Graph);~\iconseq~(Code)}
&\shortstack{---~(\iconS);~---;~---~(\iconD)}
&\shortstack{Sparse~(\embctx)~$\parallel$~Dense~(\embctx);~Sparse~(\embctx)}
&\shortstack{Generative;~LLM-based}\\
\hline
D22 & DeGPT~\cite{P146_degpt2024ndss}
&\shortstack{\iconseq~(Code)}
&\shortstack{---~(\iconD)}
&\shortstack{Sparse~(\protect\ding{55})}
&\shortstack{LLM-based}\\
 & Coda~\cite{P144_coda2019neurips}
&\shortstack{\iconseq~(Code)}
&\shortstack{Replace~(\iconS)}
&\shortstack{Sparse~(\embctx)}
&\shortstack{Generative}\\
 & BTD~\cite{P148_BTD2023usenix}
&\makecell[l]{\iconseq~(Code);~\icongraph~(TextStream);~\iconnum~(LogicExp,~TextStream)}
&\makecell[l]{---~(\iconD);~Transform;~---}
&\makecell[l]{Sparse~(\embctx);~Dense~(\protect\ding{55});~Dense~(\protect\ding{55})}
&\makecell[l]{Discriminative;~Rule-based}\\
 & \UP{IDIOMS~\cite{P161_idioms2026ndss}}
&\makecell[l]{\iconseq~(Code, Graph)}
&\makecell[l]{Replace~(\iconD)}
&\makecell[l]{Sparse~(\embctx)}
&\makecell[l]{LLM-based}\\
 & \UP{SK2Decompile~\cite{P163_sk2decompile2026iclr}}
&\makecell[l]{\iconseq~(Code)}
&\makecell[l]{---~(\iconD)}
&\makecell[l]{Sparse~(\embctx)}
&\makecell[l]{LLM-based}\\
\hline
\addlinespace[2pt]
\multicolumn{6}{@{}p{1.57\textwidth}@{}}
{
\normalsize
BinFact: binary fact;
~Code: code representation;
~Extract: subgraph extraction;
~Graph: graph representation;
~LogicExp: logical expression;
~Map: node or edge attribute mapping;
~Numeric: numerical statistic;
~Remove: filtering by removal;
~Replace: abstraction by replacement;
~Scale: feature value scaling, byte value scaling, pixel value scaling, or image size scaling;
~Transform: structural transformation
}
\end{tabular}

  }
\end{table*}

\end{document}